\documentclass[journal]{IEEEtran}
\ifCLASSINFOpdf
\else
\fi

\usepackage[utf8]{inputenc}
\usepackage[T1]{fontenc}
\usepackage{url}
\usepackage{ifthen}
\usepackage[cmex10]{amsmath} 
\usepackage{enumitem}
\usepackage{amssymb}
\usepackage{latexsym}
\usepackage{graphicx}
\usepackage{color}
\usepackage{verbatim}
\usepackage{multirow}
\usepackage{amsfonts}
\usepackage{booktabs}
\usepackage{placeins}
\usepackage{float}
\usepackage{makecell}
\usepackage{rotating}
\usepackage{calc}
\usepackage[caption=false,font=footnotesize]{subfig}

\newcommand{\R}{\ensuremath{\mathbb R}}

\newcommand{\nbu}{{\bf n}}
\newcommand{\sbu}{{\bf s}}

\newcommand{\xbu}{{\bf x}}
\newcommand{\ybu}{{\bf y}}

\newcommand{\Abu}{{\bf A}}

\newcommand{\Ibu}{{\bf I}}

\newcommand{\Pbu}{{\bf P}}

\newcommand{\Sbu}{{\bf S}}

\newcommand{\Ubu}{{\bf U}}

\newcommand{\Phibu}{{\bf \Phi}}

\newcommand{\Psibu}{{\bf \Psi}}

\newcommand{\iproof}{{\noindent \textit{Proof}}}

\newcommand{\qed}{\hfill \ensuremath{\Box}}

\newtheorem{df}{Definition}
\newtheorem{thr}{Theorem}

\newtheorem{rem}{Remark}
\newtheorem{cor}{Corollary}

\numberwithin{const2}{const}

\begin{document}
\def\smath#1{\text{\scalebox{.8}{$#1$}}}
\def\sfrac#1#2{\smath{\frac{#1}{#2}}}

%
\title{Secure and Efficient Compressed Sensing Based Encryption With Sparse Matrices}
%
%
%

\author{Wonwoo Cho,~\IEEEmembership{Student Member,~IEEE} and Nam Yul Yu,~\IEEEmembership{Senior Member,~IEEE}
\thanks{The authors are with the School of
of Electrical Engineering and Computer Science,
Gwangju Institute of Science and Technology, Gwangju 61005, South Korea.
(e-mail: ksg6604@gmail.com; nyyu@gist.ac.kr).}  
}

\maketitle

\begin{abstract}
In this paper, we study the security of a compressed sensing (CS)
based cryptosystem called a \emph{sparse one-time sensing (S-OTS)} cryptosystem,
which encrypts a plaintext with a sparse measurement matrix.
To construct the secret
matrix and renew it at each encryption,
a bipolar keystream and a random permutation pattern
are employed as cryptographic primitives,
which can be obtained by a keystream generator of stream ciphers.
With a small number of nonzero elements in the measurement matrix,
the S-OTS cryptosystem achieves efficient CS encryption
in terms of memory and computational cost.
In security analysis,
we show that the S-OTS cryptosystem can be indistinguishable
as long as each plaintext has constant energy,
which formalizes computational security against ciphertext only attacks (COA).
In addition, we consider a chosen plaintext attack (CPA) against the S-OTS cryptosystem,
which consists of two sequential stages, keystream and key recovery attacks.
Against keystream recovery under CPA,
we demonstrate that the S-OTS cryptosystem can be secure
with overwhelmingly high probability,
as an adversary needs to distinguish
a prohibitively large number of candidate keystreams.
Finally, we conduct an information-theoretic analysis to
show that the S-OTS cryptosystem can be resistant
against key recovery under CPA
by guaranteeing that the probability of success is extremely low.
In conclusion, the S-OTS cryptosystem can be
computationally secure against COA and the two-stage CPA,
while providing efficiency in CS encryption.

\end{abstract}


%
\IEEEpeerreviewmaketitle

\section{Introduction}

\IEEEPARstart{C}{ompressed} sensing (CS)~\cite{CanRomTao:robust}$-$\cite{CanTao:univ}
allows to recover a sparse signal from a much smaller number
of measurements than the signal dimension.
A signal $\xbu \in \R^N$ is called \emph{$K$-sparse}
with respect to an orthonormal sparsifying basis $\Psibu$
if ${\boldsymbol \alpha} =\Psibu \xbu$ has at most
$K$ nonzero entries, where $K \ll N$.
The sparse signal $\xbu$ is linearly measured by
$\ybu = \Phibu \xbu + \nbu = \Phibu \Psibu^T {\boldsymbol \alpha} +\nbu \in \R^M$,
where $\Phibu$ is an $M \times N$ measurement matrix with $M \ll N$
and $\nbu \in \R^M$ is the measurement noise.
In CS theory, if the sensing matrix $\Abu = \Phibu \Psibu^T$ obeys
the \emph{restricted isometry property (RIP)}~\cite{Donoho:CS}$-$\cite{Baraniuk:rip},
a stable and robust reconstruction of ${\boldsymbol \alpha}$
can be guaranteed from the incomplete measurement $\ybu$.
The CS reconstruction can be accomplished by solving
an $l_1$-minimization problem with convex optimization
or greedy algorithms~\cite{Eldar:CS}.
With efficient measurement and stable reconstruction,
the CS technique has been of interest in a variety of research fields,
e.g., communications~\cite{Tropp:beyond}$-$\cite{Haupt:toep},
sensor networks~\cite{Duarte:dist}$-$\cite{Caione:wsn},
image processing~\cite{Duarte:single}$-$\cite{Lustig:mr},
radar~\cite{Gog:radar}, etc.

The CS principle can be applied in a symmetric-key cryptosystem for
information security.
The \emph{CS-based cryptosystem} can simultaneously
compress and encrypt a plaintext $\xbu$ through a CS measurement process
by keeping the measurement matrix $\Phibu$ secret.
With the knowledge of $\Phibu$,
the ciphertext $\ybu$ can then be decrypted by
a legitimate recipient through a CS reconstruction process.
The CS-based cryptosystem can be suitable for security of
real-world applications such as multimedia,
smart grid, and the Internet of Things (IoT)~\cite{Zeng:speech}$-$\cite{Mangia:IoT},
where plaintexts of interest can be modeled to be sparse in a proper basis.
Readers are referred to~\cite{Zhang:review}
for a comprehensive review of CS in the field of information security.

Rachlin and Baron~\cite{Baron:sec} proved that
the CS-based cryptosystem cannot be perfectly secure,
but might be computationally secure.
In~\cite{Ors:secrub}, Orsdemir \emph{et al.} showed that
it is computationally secure against
a key search technique via an algebraic approach.
In~\cite{Reeves:wiretap}$-$\cite{Dau:establish},
CS-based cryptosystems have been studied in the framework of
physical layer security~\cite{Zou:wireless} by exploiting
the randomness of wireless channels.
To avoid plaintext attacks,
a CS-based cryptosystem can employ
the secret measurement matrix in a \emph{one-time sensing (OTS)} manner~\cite{Bianch:anal},
where the matrix is renewed at each encryption.
In a CS-based cryptosystem using the OTS concept,
a sender and a legitimate recipient can use
a secure random number generator (SRNG)~\cite{Menezes:crypt}
to construct the secret matrices efficiently,
by sharing only the initial seed of SRNG as a secret key.

Using random Gaussian measurement matrices in the OTS manner,
Bianchi \emph{et al.}~\cite{Bianch:anal} showed that
the \emph{Gaussian-OTS (G-OTS)} cryptosystem can be perfectly secure,
as long as each plaintext has constant energy.
In~\cite{Bianch:circ}, the authors made
a similar security analysis for a CS-based cryptosystem
which employs circulant matrices for efficient CS processes.
It has also been studied for wireless security in~\cite{Yu:Circ_SPL},
while a CS-based cryptosystem with
general partial unitary matrices embedding a keystream has been investigated in~\cite{Yu:Unitary_EU}.
In~\cite{Camb:multiclass}, Cambareri \emph{et al.} employed
random Bernoulli matrices with the OTS concept to encrypt plaintexts
sparse with respect to a non-identity orthonormal basis,
which we call the \emph{Bernoulli-OTS (B-OTS)} cryptosystem in this paper.
With the notion of asymptotic spherical secrecy,
they analyzed the security of the B-OTS cryptosystem,
asymptotically and non-asymptotically, by modeling the ciphertexts to be Gaussian distributed.
Then, they quantitatively showed that the B-OTS cryptosystem and
its class dependent variations can be resistant
against known plaintext attacks in~\cite{Camb:kpa}.
In~\cite{Cho:Wireless_CS},
the security of the asymptotically Gaussian-OTS (AG-OTS) cryptosystem,
which employs random Bernoulli matrices multiplied by a unitary matrix,
has been discussed in the presence of wireless channels.
In addition, the indistinguishability~\cite{Katz:modern}
of the G-OTS and the B-OTS cryptosystems has been studied
in~\cite{Yu:B-OTS},
which turned out to be highly sensitive to energy variation of plaintexts.

Although the OTS concept is necessary for security against plaintext attacks,
it may cause complexity issues in practical implementation.
In processing large-size signals,
renewing the measurement matrix at each encryption
would require massive data storage and computing resources.
To resolve this issue in CS imaging field,
a series of works have applied the technique of parallel CS (PCS)~\cite{Fang:PCS}
to CS-based cryptosystems in the OTS manner,
where the PCS framework can significantly reduce the size of a measurement matrix
at each encryption.
The CS-based cryptosystem proposed in~\cite{Fay:counter}
encrypts each image column-by-column,
renewing its measurement matrix at each encryption
with the counter mode of operation~\cite{Katz:modern} in block ciphers.
In this scheme, however,
the plaintexts with unequal energy result in information leakage,
which can be a cryptographic weakness.
To overcome this issue,
each plaintext should be normalized,
which requires a secure auxiliary channel
to transfer the energy information to a recipient~\cite{Hu:image}.
In~\cite{Hu:image}, 
Hu \emph{et al.} applied an additional cryptographic diffusion process
after each CS encryption of~\cite{Fay:counter} in order to prevent the information leakage.

In this paper,
we propose the \emph{sparse-OTS (S-OTS)} cryptosystem,
which employs sparse measurement matrices in the OTS manner,
to pursue efficiency and security simultaneously.
Since only a few entries of its measurement matrix
take bipolar values and all the others are zero,
the S-OTS cryptosystem can save the data storage and
reduce the computational cost required for encryption.
To renew the secret matrix at each encryption,
we employ a linear feedback shift register (LFSR) based keystream generator.
In the S-OTS cryptosystem, we show that a reliable CS decryption is
theoretically guaranteed for a legitimate recipient.

For security against ciphertext only attacks (COA),
we investigate the indistinguishability of the S-OTS cryptosystem.
If each plaintext has constant energy,
we show that the S-OTS cryptosystem can be indistinguishable,
which formalizes the notion of computational security against COA.
Then, we analyze security against chosen plaintext attacks (CPA),
which can be more threatening.
Against the S-OTS cryptosystem, this paper considers
a CPA of two sequential stages,
keystream and key recovery attacks.
At the first stage,
we verify that
the S-OTS cryptosystem can be secure against
keystream recovery under CPA with high probability,
by showing that the number of candidate keystreams is
tremendously large.
At the second stage, conducting an information-theoretic analysis,
we show that the success probability of key recovery is extremely low.

To sum up, the S-OTS cryptosystem can be computationally secure
against COA and the two-stage CPA, while providing efficient CS encryption
by using sparse measurement matrices.
Implemented in parallel,
the encryption process of the S-OTS cryptosystem can also be fast.
Although CS decryption requires high complexity
to solve an $l_1$-minimization problem,
a legitimate recipient of potential applications,
e.g., control center in IoT systems,
may have sufficiently high computing power for CS decryption.
Due to its fast and efficient encryption process, the S-OTS cryptosystem
can be a good alternative to
conventional encryption schemes, e.g., AES~\cite{Daemen:AES},
for delay-sensitive lightweight devices.

This paper is organized as follows.
First of all, Section II presents the system model,
reliability analysis, and complexity benefits of the S-OTS cryptosystem.
The indistinguishability of the S-OTS cryptosystem against COA
is investigated in Section III.
Section IV discusses adversary's CPA strategies
and the corresponding security measures.
Then, the security of the S-OTS cryptosystem
against CPA is analyzed in Section V.
Section VI numerically analyzes the security of the S-OTS cryptosystem
and demonstrates image encryption examples.
Finally, concluding remarks will be given in Section VII.

%
%
%
%

\section{System Model}

\subsection{Notations}

$u_{i,j}$, ${\bf u}^{(i)}$, ${\bf u}_j$, and $\Ubu^T$
are the entry of a matrix $\Ubu \in \R^{M \times N}$
in the $i$-th row and the $j$-th column, the $i$-th row vector, the $j$-th column vector,
and the transpose of $\Ubu$, respectively, where $1 \leq i \leq M$ and $1 \leq j \leq N$.
An identity matrix is denoted by $\Ibu$,
where its dimension is determined in the context.
For a vector $\xbu = (x_1,\cdots, x_{N})^T$,
the $l_p$-norm of $\xbu$ is denoted by
$ || \xbu ||_p = \left( \sum_{k=1} ^{N} |x_k|^p \right) ^{\frac{1}{p}}$ for $1 \le p < \infty$.
Also, $|| \xbu ||_0$ denotes the number of nonzero elements of $\xbu$.
If the context is clear,
$ || \xbu ||$ denotes the $l_2$-norm of $\xbu$.
For an index set $\Lambda$,
$|\Lambda|$ denotes the number of elements in $\Lambda$.
Finally, a vector $\nbu \sim {\mathcal {N}} ({\bf 0}, \sigma ^2 \Ibu)$
is a Gaussian random vector with mean ${\bf 0} = (0,\cdots,0)^T$
and covariance $\sigma ^2 \Ibu$.

\subsection{Sparse One-Time Sensing (S-OTS) Cryptosystem}

\subsubsection{Mathematical Formulation}

Let $\xbu \in \R^N$ be a $K$-sparse plaintext
with respect to an orthonormal sparsifying basis $\Psibu$,
i.e., $\xbu = \Psibu^T \boldsymbol{\alpha}$ with $||\boldsymbol{\alpha}||_0 \le K$.
The S-OTS cryptosystem employs a secret measurement matrix $\Phibu = \frac{1}{\sqrt{Mr}} \Sbu \Pbu$,
where $\Sbu \in \{-1,0,1\}^{M \times N}$ is a sparse matrix
containing $q$ nonzero elements in each row, $\Pbu \in \{0,1\}^{N \times N}$
is a matrix for permuting the columns of $\Sbu$,
and $r = \frac{q}{N}$ is the row-wise sparsity.
With $\Phibu$,
the S-OTS cryptosystem encrypts the plaintext $\xbu$ to provide the corresponding ciphertext
\begin{equation}\label{eq:sys_model}
\ybu = \frac{1}{\sqrt{Mr}} \Sbu \Pbu \xbu + \nbu = \frac{1}{\sqrt{Mr}} \Sbu \Pbu \Psibu^T {\boldsymbol \alpha} +\nbu,
\end{equation}
where $\nbu \sim {\mathcal {N}} ({\bf 0}, \sigma^2 \Ibu)$ is the measurement noise.
We assume $\frac{N}{M} \le q \ll \frac{N}{2}$ for efficient encryption,
where $q$ is known to an adversary.
For convenient analysis, we assume that $\eta = \frac{N}{q}$
and $Mr$ are integers throughout this paper.


\subsubsection{Keystream Generation}

To construct its secret matrix fast and efficiently in the OTS manner,
the S-OTS cryptosystem may generate nonzero elements of $\Sbu$
and a permutation pattern $\Pbu$
with SRNG.
In this paper,
we employ the \emph{self-shrinking generator (SSG)}~\cite{Meier:ssg}
to continuously generate a secure pseudorandom keystream
fast and efficiently based on LFSR.
The initial state of LFSR, or the \emph{key},
should be kept secret between a sender and a legitimate recipient,
while the structure of the keystream generator can be publicly known.
It is noteworthy that LFSR-based keystream generators
are more friendly to fast hardware implementation
than other keystream generators,
e.g., chaos-based keystream generators~\cite{Li:chaos},\cite{Zhang:chaos}.

\begin{df}\label{df:ssg}
\cite{Meier:ssg}
Assume that a $k$-stage LFSR generates a binary $m$-sequence~\cite{GolGong:SD} of
${\bf a} = (a_1,a_2,\cdots)$, where $a_i \in \{0,1\}$.
With a clock-controlled operation, the self-shrinking generator outputs $d_t=a_{2i}$
if $a_{2i-1}=1$, and discards $a_{2i}$ if $a_{2i-1}=0$.
Then, we obtain a bipolar keystream of ${\bf b} = (b_1,b_2,\cdots)$,
where $b_t=(-1)^{d_t}$ for $t=1, 2, \cdots$.
\end{df}

The SSG has a simple structure of a $k$-stage LFSR along with a clock-controlled operator.
Meier and Staffelbach~\cite{Meier:ssg} showed that the SSG keystream is
balanced, and has a period of at least $2^{\lfloor \frac{k}{2} \rfloor}$ and a linear complexity
of at least $2^{\lfloor \frac{k}{2} \rfloor -1}$.
With the nice pseudorandomness properties,
we assume that each keystream bit takes $\pm 1$ independently and uniformly at random,
which facilitates our security analysis of the S-OTS cryptosystem by
modeling the keystream bits to be truly random Bernoulli distributed.
In~\cite{Yu:AG-OTS}, numerical results demonstrated the good statistical properties
of the SSG keystream, which supports our assumption.
It is noteworthy that any other LFSR-based keystream generators can be used
to construct each measurement matrix efficiently,
as long as their keystream bits can be modeled to be Bernoulli distributed.


\begin{table}[t!]
\fontsize{8}{6pt}\selectfont
\caption{Notations and Variables}
\centering

\begin{tabular}{ll}
\toprule
{Notation}  & {Description}\\
\midrule
{$\Phibu $}                   & $M \times N$ secret measurement matrix\\[0.4\normalbaselineskip]
{$\Psibu$}                    & $N \times N$ orthonormal sparsifying basis\\[0.4\normalbaselineskip]
{$\Sbu$}                      & $M \times N$ sparse matrix embedding secret bipolar keystream\\[0.4\normalbaselineskip]
{$\Pbu$}                      & $N \times N$ secret permutation matrix\\[0.4\normalbaselineskip]
{$\Lambda_i$}                 & Index set of nonzero entries in the $i$-th row of $\Sbu$\\[0.4\normalbaselineskip]
{${\bf k}$}                   & True key of length $k$\\[0.4\normalbaselineskip]
{$\widehat{\bf k}$}           & Estimated key of length $k$\\[0.4\normalbaselineskip]
{${\bf b}^k$}                 & True consecutive keystream of length $k$\\[0.4\normalbaselineskip]
{$\widehat{\bf b}^k$}         & Estimated consecutive keystream of length $k$\\[0.4\normalbaselineskip]
{$q$}                         & Number of nonzero entries in each row of $\Phibu$, $q \ll \frac{N}{2}$\\[0.4\normalbaselineskip]
{$r$}                         & Row-wise sparsity of $\Phibu$, $r = \frac{q}{N}$\\[0.4\normalbaselineskip]
{$\rho$, $\eta$, $\tau$}      & $\rho = \frac{M}{N}$, $\eta = \frac{N}{q}$, $\tau = \lceil \frac{k}{q} \rceil$\\
\bottomrule
\end{tabular}
\label{tb:Notation}
\end{table}

\subsubsection{Secret Matrix Construction}

For given $q$ and $N$, let
\begin{equation}\label{eq:set_define}
  \Lambda_i=\{((i-1) \bmod \eta) \cdot q + l \mid l=1,\cdots,q\}
\end{equation}
be an index set of nonzero entries in the $i$-th row of $\Sbu$.
Then, $c_s=qM$ bits of the SSG keystream
are embedded in $\Sbu$, where
\begin{equation}\label{eq:s_structure}
    s_{i,j} =
    \begin{cases}
    b_{\lfloor \frac{i-1}{\eta} \rfloor \cdot N + j}, & \text{if } j \in \Lambda_i, \\
    0, & \text{otherwise}.
    \end{cases}
\end{equation}
After the S-OTS cryptosystem constructs $\Sbu$,
next $c_p$ bits of the SSG output sequence can be used to generate
a permutation pattern $\Pbu$,
where a number of algorithms that generate random permutations from coin-tossing (0 and 1)
have been studied in~\cite{Bacher:Permut} and~\cite{Bacher:MergeShuffle}.
To the best of our knowledge, $c_p \approx N \log_2 N$ on average and
the computational cost of its generation is approximately $N \log_2 N$~\cite{Bacher:MergeShuffle}.
To sum up, $\Sbu$ and $\Pbu$ can be constructed from consecutive
$c_s+c_p \approx qM + N \log_2 N$ bits of the SSG output sequence at each encryption.

\begin{table}[t!]
\fontsize{8}{6pt}\selectfont
\caption{The S-OTS Cryptosystem}
\centering

\begin{tabular}{p{1.7cm}p{6.3cm}}
\toprule
\multirow{3}{*}{Public}             & Structure of an LFSR-based keystream generator\\\cmidrule{2-2}
                                    & $\Psibu$, $q$, and $\Lambda_i$ for $i=1,\cdots,M$\\
\midrule
{Secret}                            & $\bf k$, nonzero entries of $\Sbu$, and $\Pbu$\\
\midrule
\midrule
\makecell[l]{Keystream\\generation} & \begin{minipage}{6.3cm}
                                        On input the key ${\bf k}$, the keystream generator outputs a keystream ${\bf b}$.
                                      \end{minipage}\\
\midrule
\makecell[l]{Secret matrix\\construction} & \begin{minipage}{6.3cm}
                                        On input $c_s+c_p$ bits of the keystream ${\bf b}$,
                                        the cryptosystem constructs $\Sbu$ and $\Pbu$,
                                        and then renews the keystream bits at each CS encryption.
                                      \end{minipage}\\

\midrule
{CS encryption}                     & \begin{minipage}{6.3cm}
                                        On input a plaintext $\xbu$, the cryptosystem outputs the ciphertext $\ybu = \textstyle{\sfrac{1}{\sqrt{Mr}}}\Sbu\Pbu \xbu$, where $\Sbu$ and $\Pbu$ are renewed at each encryption.
                                      \end{minipage}\\
\midrule
{CS decryption}                     & \begin{minipage}{6.3cm}
                                        On input the key ${\bf k}$ and the ciphertext $\ybu$, the plaintext $\xbu = \Psibu^T \boldsymbol{\alpha}$ is recovered by solving $\min ||\boldsymbol{\alpha}||_1 \text{ s.t. } \ybu=\Phibu \Psibu^T \boldsymbol{\alpha} + \nbu$ with the knowledge of $\Phibu$.
                                      \end{minipage}\\
\bottomrule
\end{tabular}
\label{tb:Summary}
\end{table}

A list of notations and variables, and a description of the S-OTS cryptosystem
can be found in Tables~\ref{tb:Notation} and~\ref{tb:Summary}, respectively.

\subsection{Recovery Guarantee for CS Decryption}

In CS decryption, reliability and stability
must be guaranteed for a legitimate recipient of ciphertext $\ybu$,
who knows the secret matrix $\Phibu$.
In the S-OTS cryptosystem,
the sensing matrix $\Abu = \frac{1}{\sqrt{Mr}} \Sbu \Pbu \Psibu^T$
can be interpreted as a structurally-subsampled unitary matrix~\cite{Bajwa:subsample},
as $\Psibu$ is unitary, i.e., $\Psibu \Psibu^T = \Psibu^T \Psibu = \Ibu$.
In the following, Theorem~\ref{th:rip} gives
a sufficient condition for
$\Abu$ to obey the RIP~\cite{Donoho:CS}$-$\cite{Baraniuk:rip} with high probability,
which guarantees a reliable and stable CS decryption.

\begin{thr}\label{th:rip}
\cite{Bajwa:subsample}
Let ${\mu_\Psibu} = \sqrt{N} \max_{i,j \in \{1,\cdots,N\}} |\psi_{i,j}|$ for $\Psibu$.
With positive constants $c_a$, $c_b$, $\varepsilon_1 \in (0,1)$, and $\delta_K \in (0,1)$,
the sensing matrix $\Abu = \frac{1}{\sqrt{Mr}} \Sbu \Pbu \Psibu^T$
of the S-OTS cryptosystem satisfies the RIP of order $K$
with probability exceeding $1 - 20 \max \left\{\exp \left(-c_b \frac{\delta_K^2}{\varepsilon_1^2}\right), N^{-1}\right\}$,
as long as
\begin{equation}\label{eq:rip_cond}
M \ge c_a {\mu_\Psibu^2} K \log^2 K \log^3 N \cdot \varepsilon_1^{-2}.
\end{equation}
\end{thr}

\begin{rem}\label{rem:rip_cond}
  For a legitimate recipient,
  Theorem~\ref{th:rip} shows that the S-OTS cryptosystem can theoretically
  guarantee a stable CS decryption
  with a proper choice of $\Psibu$, i.e., $\mu_\Psibu = {\mathcal O}(1)$.
  Since the sufficient condition of~\eqref{eq:rip_cond} is irrelevant to $q$,
  changing $q$ to meet the security requirement
  does not affect the recovery guarantee in CS decryption.
\end{rem}

\begin{figure}[!t]
\centering
\includegraphics[width=0.45\textwidth, angle=0]{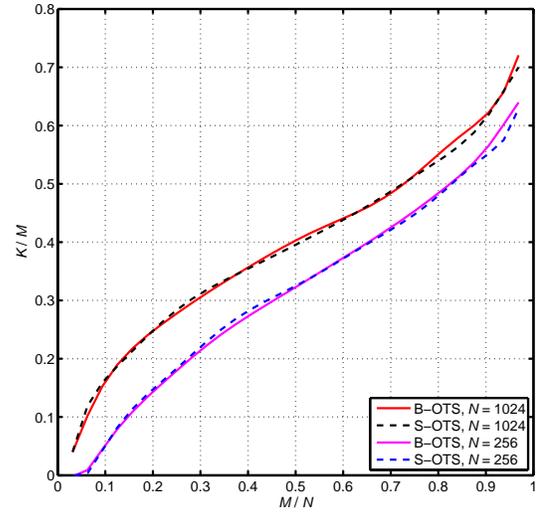}
\caption{Phase transition diagrams of
the B-OTS and the S-OTS cryptosystems,
where $q=32$ and $\Psibu$ is the DCT basis.}
\label{fig:phase}
\end{figure}

Figure~\ref{fig:phase} illustrates the phase transition diagrams
of the B-OTS and the S-OTS cryptosystems in noiseless condition, respectively,
where $\Psibu$ is the discrete cosine transform (DCT) basis and $q = 32$.
Applying the orthogonal matching pursuit (OMP)~\cite{Tropp:OMP} for CS decryption,
we tested $10^3$ different plaintexts at each test point,
where the step sizes of $\frac{M}{N}$ and $\frac{K}{M}$ are $2^{-5}$ and $10^{-2}$, respectively.
For each encryption, the plaintext $\xbu = \Psibu^T \boldsymbol{\alpha}$ is randomly generated,
where nonzero entries of $\boldsymbol{\alpha}$ are Gaussian distributed and their positions are chosen uniformly at random.
In the region below each phase transition curve,
the corresponding CS-based cryptosystem
successfully decrypts ciphertexts
with probability exceeding $99\%$,
where a decryption is declared as a success if the decrypted plaintext $\widetilde{\xbu}$ achieves
$\frac{||\xbu - \widetilde{\xbu}||^2}{||\xbu||^2}<10^{-2}$.
The figure shows that the CS decryption performance
of the S-OTS cryptosystem with $q \ll N$
is similar to that of the B-OTS cryptosystem with $q=N$ over a wide range of $M$,
which implies that the theoretical guarantee of the S-OTS cryptosystem is a bit pessimistic.

Using the S-OTS cryptosystem,
we also encrypt an $n \times n$ $8$-bit gray-scale image
in noiseless condition.
For CS encryption, all the columns of the image
are stacked into a vector $\xbu \in \R^N$, where $N=n^2$.
To examine the decryption performance
with the plaintext $\xbu$,
we employ SPGL1~\cite{Berg:SPGL1} to obtain the decrypted plaintext $\widetilde{\xbu}$
and then measure the peak signal-to-reconstruction noise ratio (PSNR)
averaged over $10^2$ different $\Phibu$,
where ${\rm PSNR} = 10 \cdot \log_{10} \left(\frac{N\cdot 255^2}{||\xbu-\widetilde{\xbu}||^2}\right)$.
To obtain $\boldsymbol{\alpha} = \Psibu\xbu$, 2D versions of the DCT,
the Daubechies 4 (D4) wavelet transform, and the Haar wavelet transform bases
are employed as $\Psibu = \Psibu_n \otimes \Psibu_n$,
where $\Psibu_n\in \R^{n\times n}$
is an 1D sparsifying basis and $\otimes$ is the Kronecker product.
Using the test image ``Lena'' with $n=256$ and $\rho = \frac{M}{N} = 0.5$,
Table~\ref{tb:RSNR_basis} shows the
average PSNR (APSNR) of a legitimate recipient
in the S-OTS cryptosystem
for various $q$ and $\Psibu$.
The S-OTS cryptosystem
guarantees a reliable CS decryption
with the DCT basis having $\mu_{\Psibu} = \mathcal{O}(1)$.
Also, we empirically found that CS decryption in the S-OTS cryptosystem
can be reliable with the D4 and the Haar wavelet bases, which
have much higher $\mu_{\Psibu} = \mathcal{O}(\sqrt{N})$.
As predicted by Remark~\ref{rem:rip_cond},
the decryption performance turns out to be irrelevant to $q$.

\subsection{Complexity Benefits}

The B-OTS cryptosystem can be computationally more efficient
than the G-OTS cryptosystem,
since the bipolar entries take less data storage
and make matrix-vector multiplications simpler.
Nevertheless, the B-OTS cryptosystem requires
$MN$ keystream bits and $MN$ operations at each encryption,
which can be a burden to lightweight systems.
By embedding fewer nonzero entries in its measurement matrix,
the S-OTS cryptosystem can reduce the number of
keystream bits and computations.
Moreover, it may have the benefit of fast encryption by conducting the matrix-vector multiplication row-wise in parallel.
Table~\ref{tb:compare} briefly compares the S-OTS and the B-OTS cryptosystems in terms of reliability and complexity.
Although they have different $M$ in theoretical recovery guarantees,
Table~\ref{tb:RSNR_basis} and Figure~\ref{fig:phase} demonstrate that
the S-OTS and the B-OTS cryptosystems empirically guarantee
similar decryption performance with the same $M$.
Furthermore, Section VI will demonstrate that the S-OTS cryptosystem
can be secure with $q \ll N$.
Thus, the S-OTS cryptosystem enjoys a significant benefit in complexity,
compared to the B-OTS cryptosystem, while guaranteeing its reliability and security.

\begin{table}[t!]
\fontsize{8}{6pt}\selectfont
\caption{CS Decryption Performance of $256 \times 256$ Lena Image ($\rho = M/N = 0.5$)}
\centering

\begin{tabular}{lcccccc}
\toprule
{CS-based}        & \multicolumn{5}{c}{\multirow{2}{*}{S-OTS}}                                             & \multirow{2}{*}{B-OTS} \\
{cryptosystem}    &                     &                     &                     &                      &                      & \\
\cmidrule{1-1}\cmidrule(lr){2-6} \cmidrule{7-7}
{No. of nonzero}  & \multirow{2}{*}{16} & \multirow{2}{*}{32} & \multirow{2}{*}{64} & \multirow{2}{*}{128} & \multirow{2}{*}{256} & \multirow{2}{*}{16384} \\
{entries in a row~($q$)}&               &                     &                     &                      &                      & \\
\midrule
{DCT}             & 29.8                & 29.7                & 29.8                & 29.6                 & 29.7                 & 29.7  \\
{D4 Wavelet}      & 32.2                & 32.3                & 32.3                & 32.4                 & 32.3                 & 32.4  \\
{Haar Wavelet}    & 30.5                & 30.5                & 30.7                & 30.6                 & 30.7                 & 30.6  \\
\bottomrule
\end{tabular}
\label{tb:RSNR_basis}
\end{table}

\begin{table}[t!]
\fontsize{8}{6pt}\selectfont
\caption{Comparison of S-OTS and B-OTS Cryptosystems}
\centering

\begin{tabular}{lll}
\toprule
{CS-based cryptosystem}                      & {S-OTS}                                           & {B-OTS}\\
\midrule
\makecell[l]{Measurements for\\recovery guarantee}    & {${\Omega}(\mu_{\Psibu}^2 K \log^2 K \log^3 N)$}  & {${\mathcal O}(K \log \frac{N}{K})$}\\ \midrule
\makecell[l]{Keystream bits \\ per encryption}        & {$qM + N \log_2 N$}                               & {$MN$}    \\\midrule
\makecell[l]{Computational cost \\ per encryption}    & {$qM + N \log_2 N$}                               & {$MN$}    \\
\bottomrule
\end{tabular}
\label{tb:compare}
\end{table}



\section{Security Analysis Against COA}

\subsection{Security Measures}

In ciphertext only attacks (COA),
an adversary tries to figure out a plaintext
by only observing the corresponding ciphertext.
We consider the \emph{indistinguishability}~\cite{Katz:modern}
to formalize the notion of computational security against COA.
In Table~\ref{tb:indist}, the \emph{indistinguishability experiment}~\cite{Katz:modern}
is described for a CS-based cryptosystem in the presence of an eavesdropper.
If no adversary passes the experiment with probability significantly better than
that of random guess, the cryptosystem is said to have the indistinguishability.
In other words, if a cryptosystem has the indistinguishability,
an adversary is unable to learn any partial information of the plaintext
in polynomial time from a given ciphertext.

\begin{table*}[t!]
\fontsize{8}{10pt}\selectfont
\caption{Indistinguishability Experiment for a CS-based Cryptosystem}
\centering
\begin{tabular}{ll}
\toprule
\emph{Step} 1: & An adversary produces a pair of plaintexts $\xbu_1$ and $\xbu_2$ of the same length, and submits them to a CS-based cryptosystem. \\
\emph{Step} 2: & The CS-based cryptosystem encrypts a plaintext $\xbu_h$ by randomly selecting $h \in \{1, 2\}$, and the corresponding ciphertext $\ybu = \Phibu \xbu_h + \nbu$\\
               & is given to the adversary. \\
\emph{Step} 3: & Given the ciphertext $\ybu$, the adversary carries out a polynomial time test  ${\mathcal D}: \ybu \rightarrow h' \in \{1, 2\}$, to figure out which plaintext was encrypted.  \\
\midrule
\emph{Decision}: & The adversary passes the experiment if $h' = h$, or fails otherwise.\\
\bottomrule
\end{tabular}
\label{tb:indist}
\end{table*}

In Table~\ref{tb:indist}, let $d_{\rm TV} (p_1, p_2)$
be the total variation (TV) distance~\cite{Gibbs:bounding}
between probability distributions
$p_1 = {\rm Pr}(\ybu | \xbu_1)$ and $p_2 = {\rm Pr}(\ybu | \xbu_2)$.
Then, it is readily checked from \cite{LeCam:asymp} that
the probability that an adversary can successfully distinguish the plaintexts 
by a binary hypothesis test ${\mathcal D}$ is bounded by
\begin{equation}\label{eq:pd}
p_d \leq \frac{1}{2} + \frac{d_{\rm TV} (p_1, p_2) }{2},
\end{equation}
where $d_{\rm TV} (p_1, p_2) \in [0, 1] $.
Therefore, if $d_{\rm TV} (p_1, p_2)$ is zero,
the probability of success
is at most that of a random guess,
which leads to the indistinguishability~\cite{Katz:modern}.

Since computing $d_{\rm TV} (p_1, p_2)$ directly is difficult~\cite{DasGupta:asymp},
we will employ an alternative distance metric to bound the TV distance.
In particular, the \emph{Hellinger} distance~\cite{Gibbs:bounding},
denoted by $d_{\rm H}(p_1, p_2)$, is useful by giving both upper and lower bounds
on the TV distance~\cite{Guntu:sharp}, i.e.,
\begin{equation}\label{eq:bnd_hell}
d_{\rm H} ^2 (p_1, p_2) \leq d_{\rm TV} (p_1, p_2) \leq d_{\rm H}(p_1, p_2) \sqrt{2 - d_{\rm H} ^2 (p_1, p_2)},
\end{equation}
where $d_{\rm H}(p_1, p_2) \in [0, 1]$.
For formal definitions and properties of the distance metrics,
see~\cite{Gibbs:bounding}$-$\cite{DasGupta:asymp}.

To analyze the security of the S-OTS cryptosystem against COA,
we examine the success probability of~\eqref{eq:pd} as a security measure.

\subsection{Indistinguishability Analysis}

In the indistinguishability experiment for the S-OTS cryptosystem,
we examine adversary's success probability
with the TV distance $d_{\rm TV}(p_1,p_2)$.
In the following, Theorem~\ref{th:tv_bnd} gives
upper and lower bounds on $d_{\rm TV}(p_1,p_2)$.

\begin{thr}\label{th:tv_bnd}
In the S-OTS cryptosystem,
let $p_1 = {\rm Pr}(\ybu | \xbu_1)$ and $p_2 = {\rm Pr}(\ybu | \xbu_2)$ in Table~\ref{tb:indist}.
For a plaintext $\xbu_h$, let ${\boldsymbol \theta}_h = \frac{\xbu_h}{||\xbu_h||}$
and $c_h = N ||{\boldsymbol \theta}_h||_4^4$ for $h=1$ and $2$, respectively.
Assuming that $\xbu_{\rm min}$ and $\xbu_{\rm max}$ are the plaintexts of minimum and maximum possible energies, respectively,
$\gamma = \frac{||\xbu_{\rm min}||^2}{||\xbu_{\rm max}||^2}$ is the minimum plaintext energy ratio
and ${\rm PNR}_{\rm max} = \frac{||\xbu_{\rm max}||^2}{M \sigma^2}$ is the maximum plaintext-to-noise power ratio (PNR)
of the cryptosystem.
Then, the worst-case lower and upper bounds on $d_{\rm TV}(p_1,p_2)$ are given by
\begin{equation*}\label{eq:tv_lowbnd}
d_{\rm TV, low} \approx 1 - \left(  \frac{4 \gamma_e}{(\gamma_e+1)^2} \right)^{\frac{M}{4}}
\cdot \left(1 - \frac{c}{8q} \left(\frac{\gamma_e-1}{\gamma_e+1} \right)^2 \right)^M,
\end{equation*}
\begin{equation*}\label{eq:tv_upbnd}
d_{\rm TV, up} \approx \sqrt{ 1 - \left(  \frac{4 \gamma_e}{(\gamma_e+1)^2} \right)^{\frac{M}{2}}
\cdot \left(1 - \frac{c}{8q} \left(\frac{\gamma_e-1}{\gamma_e+1} \right)^2 \right)^{2M}},
\end{equation*}
respectively, where
\begin{equation}\label{eq:tv_bnd_c}
c = \frac{c_{\rm max}}{(1+{\rm PNR}_{\rm max}^{-1} )^2} \cdot \left(\left(\frac{\gamma}{\gamma_e}\right)^2 +1 \right),
\end{equation}
$c_{\rm max} = \max_{\xbu_1,\xbu_2}(c_1,c_2)$, and $\gamma_e
= \frac{1+ \gamma \cdot {\rm PNR}_{\max}}{1+{\rm PNR}_{\max}}$.
\end{thr}
\iproof:
In the S-OTS cryptosystem, we can compute $d_{\rm H}(p_1,p_2)$
by replacing $N$ by $q$ in the proof of~\cite[Theorem~4]{Yu:B-OTS},
where
\begin{equation*}
d^2_{\rm H}(p_1,p_2) = 1 - \left(  \frac{4 \gamma_e}{(\gamma_e+1)^2} \right)^{\frac{M}{4}}
\left(1 - \frac{c}{8q} \left(\frac{\gamma_e-1}{\gamma_e+1} \right)^2 \right)^{M}.
\end{equation*}
Then, the lower and upper bounds on $d_{\rm TV}(p_1,p_2)$ can be given by
$d_{\rm H}(p_1,p_2)$ and \eqref{eq:bnd_hell},
which completes the proof.
\qed

\begin{cor}\label{cor:tv_bnd}
In the S-OTS cryptosystem, the success probability of an adversary
in the indistinguishability experiment is bounded by
\begin{equation}\label{eq:sp_bnd}
p_d \le \frac{1}{2}+ \frac{1}{2} \sqrt{1 -\left(  \frac{4 \gamma_e}{(\gamma_e+1)^2} \right)^{\frac{M}{2}}
\cdot \left(1 - \frac{c}{8q} \left(\frac{\gamma_e-1}{\gamma_e+1} \right)^2 \right)^{2M}}.
\end{equation}
In particular, if ${\rm PNR}_{\rm max} = \infty$ and $\gamma_e = \gamma$,
\begin{equation*}\label{eq:sp_bnd_max}
p_d \le \frac{1}{2}+ \frac{1}{2} \sqrt{ 1 - \left(  \frac{4 \gamma}{(\gamma+1)^2} \right)^{\frac{M}{2}}
\cdot \left(1 - \frac{c_{\rm max}}{4q} \left(\frac{\gamma-1}{\gamma+1} \right)^2 \right)^{2M}}.
\end{equation*}
\end{cor}

\begin{rem}\label{rem:tv_bnd}
  In~\eqref{eq:tv_bnd_c},
  $(\frac{\gamma}{\gamma_e})^2 +1 \le 2$ for $ \gamma \in (0, 1]$.
  Therefore, we need $q \ge \frac{c_{\rm max}}{4 \cdot (1+{\rm PNR}_{\rm max}^{-1} )^2}$
  to guarantee $\frac{c}{8q} \le 1$ for all possible $\gamma$,
  which makes the upper bound of~\eqref{eq:sp_bnd} valid
  when ${\rm PNR}_{\max}$ is given.
  In addition,
  $q \ge \frac{c_{\rm max}}{4}$ ensures
  that the bound is valid for any ${\rm PNR}_{\max}$.
  As a result, the S-OTS cryptosystem can be indistinguishable
  with any choice of $q \ge \frac{c_{\rm max}}{4}$,
  as long as each plaintext has constant energy, i.e., $\gamma=1$.
  Note that the indistinguishability can be achieved \emph{asymptotically},
  due to the Gaussian approximation~\cite[Remark~2]{Yu:B-OTS}.
\end{rem}

\begin{rem}\label{rem:tv_conv}
The upper bound of $p_d$ in~\eqref{eq:sp_bnd} converges to $\frac{1}{2}$ as $\gamma$ goes to $1$,
where the convergence speed depends on $\frac{c_{\rm max}}{q}$
for given ${\rm PNR}_{\rm max}$.
To achieve faster convergence, it is necessary to have
lower $\frac{c_{\rm max}}{q}$ in the S-OTS cryptosystem.
We can obtain lower $c_{\rm max}$ if the energy of $\xbu$ is distributed
as uniformly as possible in each element~\cite{Yu:B-OTS}.
When $c_{\rm max}$ is given,
we need to increase $q$ to obtain a lower $\frac{c_{\rm max}}{q}$.
\end{rem}




\section{CPA Against the S-OTS Cryptosystem}

\subsection{Two-stage CPA}

Chosen plaintext attacks (CPA)
against the S-OTS cryptosystem aim to retrieve the initial state of SSG,
or the key ${\bf k}$ of length $k$,
from the pairs of a deliberately chosen plaintext and the corresponding ciphertext.
However, the SSG has a remarkable resistance against
known cryptanalytic attacks~\cite{Mihalj:faster}$-$\cite{Zhang:Guess},
even if a consecutive SSG keystream is observed.
Since an adversary needs to make further efforts to reconstruct ${\bf k}$
after restoring a consecutive SSG keystream,
this paper considers a CPA of two sequential stages against the S-OTS cryptosystem,
keystream and key recovery attacks.

To retrieve ${\bf k}$ at the second stage,
an adversary may deploy the key search algorithm in~\cite{Zhang:Guess}
requiring ${\mathcal O}(2^{0.161 k})$ consecutive SSG keystream bits,
where the search complexity is ${\mathcal O}(2^{0.556 k})$.
However, the probability to successfully recover such a long keystream at the first stage
will be extremely low,
since the keystream can be placed across multiple encryptions
with different permutation patterns.
Thus, we assume that an adversary tries
to observe a short keystream of length ${\mathcal O}(k)$ at the first stage,
and then employs
the key search algorithms in~\cite{Zenner:imp}$-$\cite{Hell:TNA} at the second stage
with the search complexity of ${\mathcal O}(2^{\lambda k})$, where $\lambda \in (0,1)$.
To the best of our knowledge, $\lambda_{\rm min} \approx 0.66$ from~\cite{Hell:TNA},
where $\lambda_{\rm min}$ is the smallest $\lambda$ among the algorithms.

To analyze the security of the S-OTS cryptosystem
against the two-stage CPA, we impose some mild assumptions without loss of generality.

\begin{enumerate}
  \item[A1)]The measurement noise is negligibly small,
  so noiseless ciphertexts are available for an adversary.
  \item[A2)] In cryptanalysis, an adversary with bounded computing power cannot
  execute any detection algorithm with complexity greater than $2^L$,
  where $L>0$.
  \item[A3)] The key length is sufficiently large,
  i.e., $k>L$, which makes a brute-force key search infeasible.
  \item[A4)] To deploy key search algorithms,
  it is sufficient for an adversary to observe $k$ consecutive keystream bits.
  \item[A5)] In $\Phibu$, $N \ge k$, which allows an adversary
  to recover $k$ consecutive keystream bits
  from a single measurement matrix by guaranteeing
  $qM \ge N \ge k$.
\end{enumerate}

Under the assumptions,
Table~\ref{tb:cpa} describes the two-stage CPA against the S-OTS cryptosystem,
which exploits a single plaintext-ciphertext pair.
At the first stage (Step 3 of Table~\ref{tb:cpa}),
an adversary tries to recover a true keystream of length $k$,
or ${\bf b}^k$, which is hidden during encryption,
by solving equations with respect to the plaintext-ciphertext pair.
The adversary then attempts to deduce the true key ${\bf k}$
from the estimated keystream $\widehat{\bf b}^k$
at the second stage (Step 4 of Table~\ref{tb:cpa}).

\begin{table*}[t!]
\fontsize{8}{10pt}\selectfont
\caption{Two-stage CPA Against the S-OTS Cryptosystem}
\centering
\begin{tabular}{ll}
\toprule
\emph{Step} 1:   & An adversary produces a plaintext $\xbu$ and submit it to the S-OTS cryptosystem. \\
\emph{Step} 2:   & The S-OTS cryptosystem encrypts the plaintext $\xbu$ and gives the corresponding ciphertext $\ybu = \Phibu \xbu$ back to the adversary. \\
\emph{Step} 3:   & At the first stage, the adversary attempts to recover a consecutive keystream ${\bf b}^k$ by solving~\eqref{eq:plain_row} from the plaintext-ciphertext pair $(\xbu, \ybu)$,\\
                 & which yields an estimated keystream $\widehat{\bf b}^k$.\\
\emph{Step} 4:   & At the second stage, the adversary attempts to recover the key ${\bf k}$ using $\widehat{\bf b}^k$, which yields an estimated key $\widehat{\bf k}$.\\
\midrule
\emph{Decision}: & The adversary's two-stage CPA succeeds if $\widehat{\bf k} = {\bf k}$, or fails otherwise.\\
\bottomrule
\end{tabular}
\label{tb:cpa}
\end{table*}

\subsection{Stage 1: Keystream Recovery Attacks}
Since $\Sbu$ has $q$ consecutive keystream bits in each row,
a true consecutive SSG keystream ${\bf b}^k$
can be obtained by recovering $\tau = \lceil \frac{k}{q} \rceil$
consecutive rows of $\Sbu$.
Therefore, an adversary can attempt to obtain an estimated keystream $\widehat{\bf b}^k$
satisfying
\begin{equation}\label{eq:plain_row}
  y_i = \sum_{j =1}^N \phi_{i,j} x_j
\end{equation}
for $i=1,\cdots,\tau$.
In the S-OTS cryptosystem, the permutation pattern of $\Pbu$
requires additional complexity for an adversary to reconstruct ${\bf b}^k$ from~\eqref{eq:plain_row},
by diffusing a consecutive keystream across $\Phibu$.
To obtain $\widehat{\bf b}^k$ in the presence of $\Pbu$,
the plaintexts of an adversary's choice
can be classified into two classes.
\setlist[description]{font=\normalfont\space}
\begin{description}[leftmargin=!,labelwidth=\widthof{$-$ }]
  \item[$-$] The first class includes plaintexts each of which enables
  an adversary to bypass $\Pbu$ by $\Pbu \xbu = \xbu$.
  Obviously, the plaintexts of this class have the form of
  $\xbu=(a,\cdots,a)^T$ with a nonzero constant $a$.
  \item[$-$] The second class contains plaintexts each of
  which recovers all the entries of $\Phibu$ by a single CPA.
  In particular, if $\xbu = (2^0,2^1,\cdots,2^{N-1})^T$,
  all the entries of $\Phibu$ can be restored, while $\Sbu$ and $\Pbu$ are unknown.
\end{description}
We believe that bypassing $\Pbu$ or recovering $\Phibu$, the two classes
may require lower complexity for keystream recovery under CPA than any other selection of plaintext.
Thus, we investigate the security of the S-OTS cryptosystem by
applying plaintexts chosen from the above two classes.
A security analysis employing a more efficient 
selection of plaintext is left open for future research.

In the first class,
if an adversary applies a plaintext $\xbu=(a,\cdots,a)^T$
 with $a = \sqrt{Mr}$,
\eqref{eq:plain_row} becomes
\begin{equation}\label{eq:ML_model}
y_i = \sum_{j \in \Lambda_i} s_{i,j}
\end{equation}
for $i=1,\cdots,\tau$, where $\Lambda_i$ is defined by~\eqref{eq:set_define}
and $s_{i,j}$ takes $\pm 1$ for $j \in \Lambda_i$.
Let $q_i^+ = |\Lambda_i^+|$ and $q_i^- = |\Lambda_i^-|$, respectively,
where $\Lambda_i^+ = \{j \mid s_{i,j}=+1, j\in \Lambda_i\}$ and
$\Lambda_i^- = \{j \mid s_{i,j}=-1, j\in \Lambda_i\}$.
Then, $q_i^+ = \frac{1}{2}(q+y_i)$
and $q_i^- = \frac{1}{2}(q-y_i)$ from~\eqref{eq:ML_model},
where the adversary obtains the numbers of $+1$'s and $-1$'s
in the $i$-th row of $\Sbu$.

With a plaintext $\xbu = (2^0,2^1,\cdots,2^{N-1})^T$,
all the entries of a secret measurement matrix in the B-OTS cryptosystem
can be easily recovered by a single CPA~\cite{Zhang:bilevel}.
Similarly, by choosing the plaintext $\xbu$ in the second class,
an adversary can successfully restore all the entries of
$\Phibu = \frac{1}{\sqrt{Mr}} \Sbu \Pbu$ in the S-OTS cryptosystem.
If $N \ge k$ under A5),
the adversary can obtain $q_i ^+$ and $q_i ^-$
for $i=1,\cdots, \tau$, from the entries of $\Phibu$,
but with no perfect knowledge of $\Pbu$.

\begin{rem}\label{rem:permut}
  From an attack with the second class plaintext,
  an adversary can reduce the number of possible candidates of $\Pbu$
  by observing the positions of nonzero entries of $\Phibu$,
  which is not possible by an attack with the first class plaintext.
  Also, if $\Pbu$ is restored completely, the keystream bits embedded in
  $\Sbu$ can be directly obtained from $\Phibu$ and $\Pbu$.
  However, we assume that an adversary makes no attempt to
  recover $\Pbu$
  in the attack with the second class plaintext,
  since retrieving $\Pbu$
  may still require an extremely large number of computations,
  as shown in an example attack of Appendix~\ref{pf:permut_keystream}.
  To analyze the effect of $\Pbu$ on the security of the S-OTS cryptosystem thoroughly, a further research will be necessary.
\end{rem}

In summary,
this paper assumes that an adversary trying to recover ${\bf b}^k$ exploits
the information of $q_i^+$ and $q_i^-$ for $i=1,\cdots,\tau$, which can be obtained by applying either of the two classes of plaintexts.
Thus, we count the number of $\widehat{\bf b}^k$
satisfying~\eqref{eq:ML_model} for $i=1,\cdots,\tau$,
as a security measure against keystream recovery under CPA,
which can be applicable to the attacks with both classes of plaintexts.

\subsection{Stage 2: Key Recovery Attacks}

After estimating a consecutive SSG keystream $\widehat{\bf b}^k$,
an adversary tries to retrieve the key ${\bf k}$ of the S-OTS cryptosystem.
Once a true SSG keystream ${\bf b}^k$ has been successfully recovered,
or $\widehat{\bf b}^k = {\bf b}^k$,
the adversary is able to reconstruct ${\bf k}$
via the key search algorithms in~\cite{Zenner:imp}$-$\cite{Hell:TNA},
as long as $\lambda_{\rm min} k \le L$.
If $\lambda_{\rm min} k > L$ from a sufficiently long key,
we assume that no adversary is able to exploit such key search algorithms,
even when $\widehat{\bf b}^k = {\bf b}^k$.
Then, we conduct an information-theoretic analysis
to develop an upper bound on the success probability of key recovery
${\rm P}_{\rm key} = {\rm Pr}[\widehat{\bf k} = {\bf k}]$,
which will be used
as a security measure against key recovery under CPA.

\section{Security Analysis Against CPA}

\subsection{Stage 1: Keystream Recovery Attacks}

At the first stage of CPA,
we assumed in Section IV.B that an adversary
attempts to obtain an estimated keystream $\widehat{\bf b}^k$ satisfying~\eqref{eq:ML_model} for $i=1,\cdots,\tau$,
by exploiting the numbers of $+1$'s and $-1$'s
in each row of $\Sbu$.
In what follows, Theorem~\ref{th:cpa_sol} gives a lower bound
on the number of possible $\widehat{\bf b}^k$.

\begin{thr}\label{th:cpa_sol}
Let $\mathcal{S}_{\rm CPA}$ be the number of possible keystreams of length $k$,
when an adversary attempts to
reconstruct a true keystream ${\bf b}^k$
with the knowledge of the numbers of $+1$'s and $-1$'s in each row of $\Sbu$.
Then,
\begin{equation}\label{eq:CPA_sol}
\mathcal{S}_{\rm CPA} \ge \binom{q}{\left\lceil\frac{q-t}{2}\right\rceil}^\tau \triangleq {\mathcal S}_{\rm CPA, low}
\end{equation}
with probability exceeding $1-\varepsilon_2$ for small $\varepsilon_2 \in (0,1)$,
where $\tau = \lceil \frac{k}{q} \rceil$ and
$t=\sqrt{2q \cdot \log \frac{2}{1-(1-\varepsilon_2)^{\frac{1}{\tau}}}} \in [0,q]$.
\end{thr}

\iproof: See Appendix~\ref{pf:cpa_sol}.

In Theorem~\ref{th:cpa_sol}, if $k$ and $\varepsilon_2$ are fixed,
${\mathcal S}_{\rm CPA, low}$ is irrelevant to $N$ and only depends on $q$.
Therefore, we can easily adjust the lower bound of \eqref{eq:CPA_sol} by changing $q$.
In the two-stage CPA,
if $\mathcal{S}_{\rm CPA} > 2^L$ with a large $q$,
the intractability of keystream recovery
at the first stage prohibits
an adversary from finding the true key at the second stage.
In what follows,
Theorem~\ref{th:cpa_q} gives a sufficient condition on $q$
to guarantee $\mathcal{S}_{\rm CPA} > 2^L$
with high probability,
which makes keystream recovery under CPA infeasible.

\begin{thr}\label{th:cpa_q}
The S-OTS cryptosystem guarantees
$\mathcal{S}_{\rm CPA} > 2^L$
with probability exceeding $1-\varepsilon_2$,
if $k \ge L \cdot e \log 2$ and
\begin{equation}\label{eq:CPA_p}
q \ge \frac{1}{2} \left(2+\frac{4}{\beta - 2}\right)^2 \log \frac{2}{1-(1-\varepsilon_2)^{\frac{1}{k\rho +1}}} \triangleq q_{\rm CPA},
\end{equation}
where $\rho = \frac{M}{N}$,
$\beta =  -\frac{k}{L \log 2} {\mathcal W}_{-1} \left(-\frac{L \log 2}{k}\right)$,
and $\mathcal{W}_{-1} (\cdot)$ is the lower branch of
Lambert \emph{W} function~\cite{Corless:lamW}.
\end{thr}

\iproof: See Appendix~\ref{pf:CPA_q}.

Theorem~\ref{th:cpa_q} demonstrates that if each row of $\Phibu$
takes more nonzero entries than $q_{\rm CPA}$,
keystream recovery under CPA is theoretically infeasible with high probability.
In what follows,
Corollary~\ref{cor:cpa_q} gives the largest possible value of $q_{\rm CPA}$.

\begin{cor}\label{cor:cpa_q}
In Theorem~\ref{th:cpa_q}, if $k \ge L \cdot e \log 2$,
then $\beta \ge e$, which leads to
\begin{equation}\label{eq:CPA_p_2}
q_{\rm CPA} \le \frac{1}{2} \left(2+\frac{4}{e - 2}\right)^2 \log \frac{2}{1-(1-\varepsilon_2)^{\frac{1}{k\rho +1}}} \triangleq q_{\rm CPA,up}.
\end{equation}
\end{cor}

Corollary~\ref{cor:cpa_q} implies that
if we choose $q > q_{\rm CPA,up}$, then
$\mathcal{S}_{\rm CPA} > 2^L$ with high probability
for every $L \le \frac{k}{e \log 2}$.
By Theorem~\ref{th:cpa_q} and Corollary~\ref{cor:cpa_q},
the security parameter $q$ should be as large as possible to ensure that
the S-OTS cryptosystem can be secure against keystream recovery under CPA.

If the S-OTS cryptosystem has $\mathcal{S}_{\rm CPA} \le 2^L$,
an adversary may be able to obtain a true keystream ${\bf b}^k$.
In Corollary~\ref{cor:cpa_prob}, we derive
an upper bound on the probability that the adversary
successfully recovers ${\bf b}^k$ with its computing power,
where the proof is straightforward from~\eqref{eq:cpa_prob} and Appendix~\ref{pf:CPA_q}.

\begin{cor}\label{cor:cpa_prob}
Let ${\rm P}_{\rm suc} = {\rm Pr}\left[\mathcal{S}_{\rm CPA} \le 2^L\right]$ be the probability that an adversary may succeed in keystream recovery under CPA
with the bounded computing power of $2^L$.
If the S-OTS cryptosystem satisfies $k \ge L \cdot e \log 2$,
we have
\begin{equation*}
{\rm P}_{\rm suc} \le 1- \left(1-2e^{-\frac{q}{2}\left(1-\frac{2}{\beta}\right)^2}\right)^\tau \triangleq {\rm P}_{\rm suc,up},
\end{equation*}
where $\tau = \lceil \frac{k}{q} \rceil$ and
$\beta =  -\frac{k}{L \log 2} {\mathcal W}_{-1}\left(-\frac{L \log 2}{k}\right)$.
\end{cor}

\begin{rem}
As long as $k \ge L \cdot e \log 2$, we have $\beta \ge e$, which implies that
${\rm P}_{\rm suc,up} < \tau \cdot 2e^{-\frac{q}{2}\left(1-\frac{2}{e}\right)^2}$
in Corollary~\ref{cor:cpa_prob}.
Note that the bound exponentially decays as $q$ grows larger.
Thus, if $k \ge L \cdot e \log 2$, the success probability of keystream recovery
under CPA disappears exponentially over $q$,
and is \emph{negligible}~\cite{Katz:modern}, regardless of $L$.
In summary, the S-OTS cryptosystem is computationally secure
against keystream recovery under CPA in an asymptotic manner.
\end{rem}

\subsection{Stage 2: Key Recovery Attacks}

At the second stage of CPA against the S-OTS cryptosystem,
an adversary attempts to recover
the key ${\bf k}$ from an estimated keystream $\widehat{\bf b}^k$,
which is obtained at the first stage.
If $\widehat{\bf b}^k = {\bf b}^k$,
the adversary can successfully reconstruct ${\bf k}$
via the key search algorithms in~\cite{Zenner:imp}$-$\cite{Hell:TNA}
with the complexity of $\mathcal{O}(2^{\lambda k})$,
as long as $\lambda_{\rm min} k \le L$.
With a long key satisfying $\lambda_{\rm min} k > L$,
no known key search algorithms for SSG can succeed
even when $\widehat{\bf b}^k = {\bf b}^k$.
Note that $k \ge L \cdot e \log 2$ in
Theorem~\ref{th:cpa_q} is sufficient
to achieve $\lambda k > L$ for all
$\lambda \ge \lambda_{\rm min} = 0.66$ in~\cite{Hell:TNA}.
Alternatively, we employ an information-theoretic tool to investigate
the key recovery performance for the S-OTS cryptosystem.
Under the condition that an adversary has obtained $\widehat{\bf b}^k$
at the first stage,
we consider a hypothesis testing
that the adversary chooses a candidate key $\widehat{\bf k}$
among $2^{\delta k}$ hypotheses, where $\frac{1}{k} \le \delta \le 1$.
In what follows,
Theorem~\ref{th:key_prob} gives an upper bound
of the success probability of key recovery, or
${\rm P}_{\rm key} = {\rm Pr}[\widehat{\bf k} = {\bf k}]$.

\begin{thr}\label{th:key_prob}
Based on $\widehat{\bf b}^k$,
an adversary chooses a candidate key $\widehat{\bf k}$
among $2^{\delta k}$ hypotheses, where $\frac{1}{k} \le \delta \le 1$.
Let ${\rm P}_{\rm key}$ be the probability
that an adversary successfully recovers the key of the S-OTS cryptosystem
at the second stage of CPA. Then,
\begin{equation}\label{eq:key_prob}
{\rm P}_{\rm key} \le 2^{-k}+\left(1-2^{-k}-\delta + \frac{1}{k}\right) \cdot {\rm P}_{\rm suc,up} \triangleq {\rm P}_{\rm key, up},
\end{equation}
where ${\rm P}_{\rm suc,up}$ is the upper bound on the success probability
of keystream recovery under CPA in Corollary~\ref{cor:cpa_prob}.
\end{thr}

\iproof:
See Appendix~\ref{pf:key_prob}.

\begin{rem}\label{rem:key_prob}
In~\eqref{eq:key_prob},
${\rm P}_{\rm key, up}$ only depends on ${\rm P}_{\rm suc, up}$
when $k$ and $\delta$ are given.
Therefore,
reducing the success probability of keystream recovery with a proper $q$ can
yield a low success probability of key recovery,
which leads to the security of the S-OTS cryptosystem against the two-stage CPA.
\end{rem}

\subsection{Key Refresh Time}

When an adversary attempts the two-stage CPA repeatedly,
one can renew the key in every ${\rm T}_{\rm ref}$ encryptions
to keep the S-OTS cryptosystem secure,
where ${\rm T}_{\rm ref}$ is the \emph{key refresh time}.
In what follows,
Theorem~\ref{th:keyrefresh} provides a sufficient condition on ${\rm T}_{\rm ref}$
to guarantee security against the two-stage CPA with high probability.

\begin{thr}\label{th:keyrefresh}
For small $\varepsilon_3 \in (0,1)$,
the S-OTS cryptosystem guarantees security against the two-stage CPA
within ${\rm T}_{\rm ref}$ encryptions
with probability exceeding $1-\varepsilon_3$, if
\begin{equation}\label{eq:keyref}
{\rm T}_{\rm ref} \le \frac{\log (1-\varepsilon_3)}{\log (1-{\rm P}_{\rm key,up})} \triangleq {\rm T}_{\rm ref,up},
\end{equation}
where ${\rm P}_{\rm key,up}$ is the upper bound
on the success probability of key recovery under CPA in Theorem~\ref{th:key_prob}.
\end{thr}

\iproof:
Since the S-OTS cryptosystem renews its measurement matrix at each CS encryption,
the probability that the S-OTS cryptosystem is secure against
${\rm T}_{\rm ref}$ repeated CPA is given by $(1-{\rm P}_{\rm key})^{{\rm T}_{\rm ref}}$,
which yields~\eqref{eq:keyref} immediately from~\eqref{eq:key_prob}.
\qed

In the proof of Theorem~\ref{th:keyrefresh},
we assumed that an adversary has no benefits by applying
multiple plaintext-ciphertext pairs to the S-OTS cryptosystem,
due to the usage of keystreams in a one-time manner.
However, this assumption does not take into account the potential of
a more elaborate strategy of an adversary.
More research efforts will be necessary to analyze the security
of the S-OTS cryptosystem against
repeated CPA with multiple ciphertext-plaintext pairs, which is left open.

As the SSG keystream has a period of at least $2^{\lfloor \frac{k}{2} \rfloor}$~\cite{Meier:ssg},
note that
$(c_s + c_p) \cdot {\rm T}_{\rm ref} < 2^{\lfloor \frac{k}{2} \rfloor}$ must be satisfied
to prevent reuse of keystream bits,
where $c_s$ and $c_p$ are the lengths of SSG output sequences
required to construct $\Sbu$ and $\Pbu$ in each encryption, respectively.

\section{Numerical results}

This section presents numerical results
of the indistinguishability and the security against the two-stage CPA of the S-OTS cryptosystem.
Also, we demonstrate digital image encryption examples.

\subsection{Indistinguishability}

Table~\ref{tb:basis_c} shows the empirical values of $c_{\rm max}$
in Theorem~\ref{th:tv_bnd} for various $\Psibu$ and $N$.
Each $c_{\rm max}$ is measured over $10^6$ plaintext pairs
$\xbu = \Psibu^T {\boldsymbol \alpha}$,
where ${\boldsymbol \alpha}$ has Gaussian distributed nonzero entries with $K=8$.
We can notice that the DCT
and the Walsh-Hadamard transform (WHT) bases,
which have no zero entries,
yield low $c_{\rm max}$ with dense $\xbu$.
On the other hand, the D4 and the Haar wavelet transform bases,
which have many zero entries, yield high $c_{\rm max}$.
According to Remarks~\ref{rem:tv_bnd} and~\ref{rem:tv_conv},
one needs $\frac{c_{\rm max}}{q} \le 4$ for a valid upper bound of $p_d$,
which should be as low as possible for fast convergence of the bound.
Therefore, the S-OTS cryptosystem may require a large $q$
when we employ a basis $\Psibu$ with high $c_{\rm max}$,
like the D4 and the Haar wavelet bases,
which compromises its efficiency.

\begin{table}[t!]
\fontsize{8}{6pt}\selectfont
\caption{Empirical $c_{\rm max}$ with Different Sparsifying Basis and $N$}
\centering

\begin{tabular}{lcccccc}
\toprule
{$N$}  & {$64$} & {$128$} & {$256$} & {$512$} & {$1024$} \\
\midrule
{DCT}                    & 5.6               & 4.8                 & 4.5                 & 4.2                  & 4.0                 \\
{WHT}                    & 5.0               & 4.6                 & 4.4                 & 4.1                  & 4.0                 \\
{D4 Wavelet}             & 50.0              & 89.4                & 186.8               & 357.5                & 684.8                 \\
{Haar Wavelet}           & 49.7              & 82.5                & 163.9               & 319.8                & 555.4                 \\
\bottomrule
\end{tabular}
\label{tb:basis_c}
\end{table}

\begin{figure}[!t]
\centering
\includegraphics[width=0.45\textwidth, angle=0]{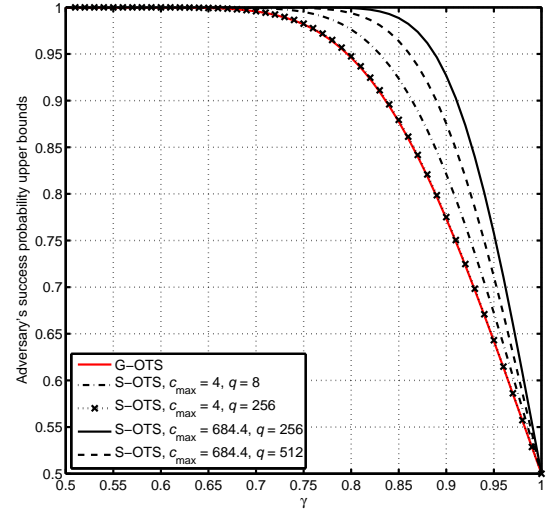}
\caption{Upper bounds of $p_d$ over $\gamma$ in the noiseless
S-OTS and G-OTS cryptosystems with various $q$ and $c_{\rm max}$,
where $N=1024$, $M=256$,
$c_{\rm max} = 4$ for the DCT or the WHT basis, and $c_{\rm max} = 684.4$ for the D4 wavelet transform basis.}
\label{fig:noiseless_indist_comp}
\end{figure}

Figure~\ref{fig:noiseless_indist_comp} depicts the upper bounds of
the success probability of an adversary in the indistinguishability experiment over $\gamma$
for the S-OTS and the G-OTS cryptosystems, respectively,
where $N=1024$, $M=256$, and ${\rm PNR}_{\rm max} = \infty$.
Even though the upper bound of~\eqref{eq:sp_bnd}
cannot be smaller than that of the G-OTS cryptosystem
presented in~\cite[Corollary 1]{Yu:B-OTS},
the figure implies that we can make them closer to each other
with lower $\frac{c_{\rm max}}{q}$.
If $c_{\rm max}=4$ for the DCT or the WHT basis,
numerical results revealed that
the difference of the upper bounds between the S-OTS and the G-OTS cryptosystems
is less than $10^{-2}$ for $q \ge 48$.
If $c_{\rm max}=684.4$ for the D4 wavelet transform basis,
it is necessary to have $q \ge 172$ to make the upper bound valid.
In this case, the figure demonstrates that
the S-OTS cryptosystem with such a high $c_{\rm max}$ cannot make
the upper bound close to that of the G-OTS cryptosystem even with $q = \frac{N}{2}$.

\subsection{Security Against CPA}

\begin{figure}[!t]
\centering
\includegraphics[width=0.45\textwidth, angle=0]{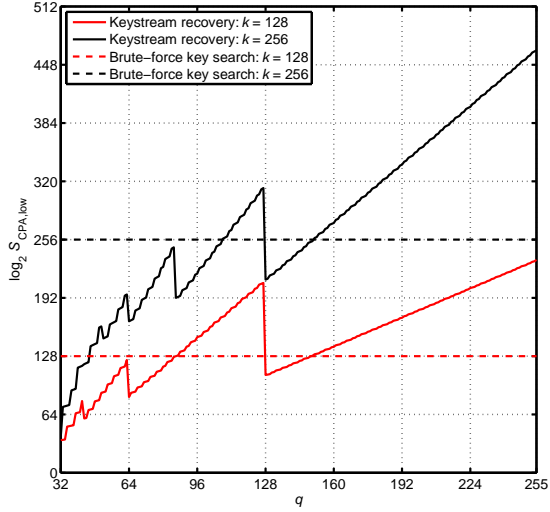}
\caption{$\log_2 {\mathcal S}_{\rm CPA,low}$ over $q$
for $k=128$ and $256$, where $\varepsilon_2$ = $10^{-5}$.}
\label{fig:sol_q}
\end{figure}

Figure~\ref{fig:sol_q} sketches $\log_2 {\mathcal S}_{\rm CPA, low}$
of Theorem~\ref{th:cpa_sol} over $q$ for $k=128$ and $256$, where $\varepsilon_2 = 10^{-5}$.
The figure shows that ${\mathcal S}_{\rm CPA, low}$
does not monotonically increase over $q$,
but drops whenever $\tau= \lceil \frac{k}{q} \rceil$ changes its value.
Thus, one needs to choose $q$ carefully,
to avoid such drops and get a higher ${\mathcal S}_{\rm CPA, low}$.
Moreover, if a selection of $q$ yields
$\log_2 {\mathcal S}_{\rm CPA, low} > k$,
the keystream recovery attack can be computationally more expensive than
a brute-force key search.
For example,
if $k=256$ and $q \in  \{108,\cdots, 127, 151,\cdots, 255, 279,\cdots\}$,
$\mathcal{S}_{\rm CPA}> 2^k$ with probability exceeding $1-10^{-5}$.
With such $q$, a brute-force key search would be a better strategy,
which demonstrates the security of the S-OTS cryptosystem against keystream recovery under CPA.


\begin{figure}[!t]
\centering
\includegraphics[width=0.45\textwidth, angle=0]{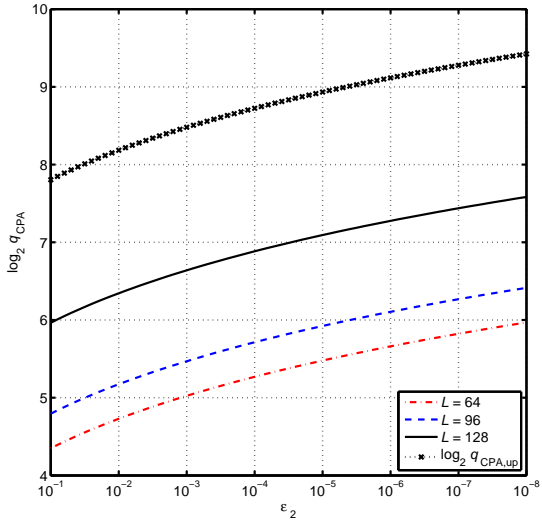}
\caption{$\log_2 {q}_{\rm CPA}$
and $\log_2 {q}_{\rm CPA,up}$ over $\varepsilon_2$ with various $L$, where $k=256$ and $\rho=0.5$.}
\label{fig:q_CPA}
\end{figure}

Figure~\ref{fig:q_CPA} depicts $\log_2 q_{\rm CPA}$ of Theorem~\ref{th:cpa_q}
over $\varepsilon_2$ for $\rho = 0.5$, $k=256$, and $L \le 128$,
where $k \ge L \cdot e \log 2$ is met.
The figure shows that
if $q \ge 137$, then $\mathcal{S}_{\rm CPA} > 2^{128}$ with probability exceeding $1-10^{-5}$,
which suggests that the S-OTS cryptosystem with such $q$ can be secure
against keystream recovery under CPA from an adversary
with computing power of at most $2^{128}$.
In addition,
$q_{\rm CPA, up} < 512$ at $\varepsilon_2 = 10^{-5}$, which implies that $q \ge 512$
ensures $\mathcal{S}_{\rm CPA} > 2^L$,
or the infeasibility of keystream recovery under CPA,
for any $L < \frac{k}{e \log 2}$
with probability exceeding $1-10^{-5}$.

\begin{figure}[!t]
\centering
\includegraphics[width=0.45\textwidth, angle=0]{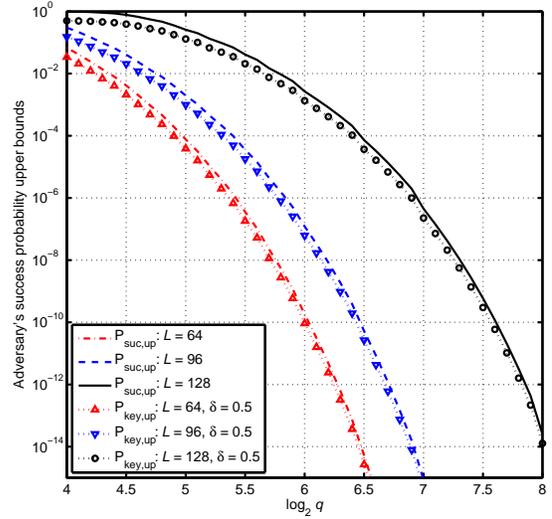}
\caption{${\rm P}_{\rm suc,up}$
and ${\rm P}_{\rm key,up}$ over $\log_2 q$ with various $L$, where $k=256$.}
\label{fig:prob_q}
\end{figure}

\begin{figure}[!t]
\centering
\includegraphics[width=0.45\textwidth, angle=0]{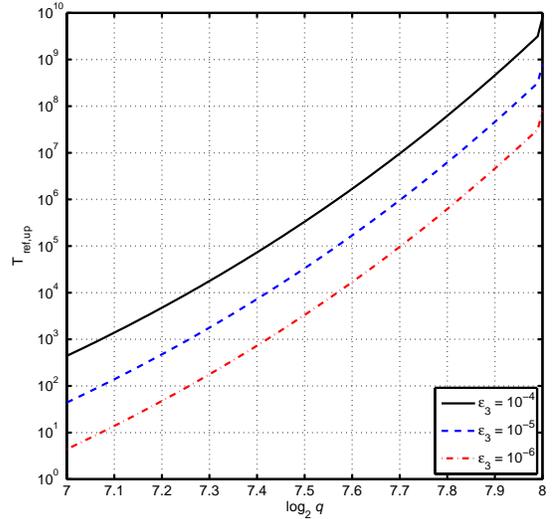}
\caption{${\rm T}_{\rm ref,up}$ over $\log_2 q$ with various $\varepsilon_3$, where $\delta=0.5$.}
\label{fig:keyrefresh}
\end{figure}

Figure~\ref{fig:prob_q} sketches ${\rm P}_{\rm suc,up}$ of Corollary~\ref{cor:cpa_prob}
and ${\rm P}_{\rm key,up}$ of Theorem~\ref{th:key_prob} over $\log_2 q$
to demonstrate the security of the S-OTS cryptosystem against the two-stage CPA,
where $\delta = 0.5$.
When $k=256$ and $L \le 128$, the figure shows that
$q \ge 128$ guarantees ${\rm P}_{\rm suc} < 10^{-6}$.
In addition, Figure~\ref{fig:keyrefresh} depicts
the key refresh time ${\rm T}_{\rm ref}$ over $\log_2 q$
with $\delta = 0.5$.
When $q = 256$ and $\varepsilon_3 = 10^{-5}$,
we have ${\rm T}_{\rm ref}>10^8$,
which implies that the S-OTS cryptosystem can be secure against the two-stage CPA
using the same key for $10^8$ encryptions
by keeping ${\rm P}_{\rm key} < 10^{-5}$.

To sum up,
Figures~\ref{fig:sol_q}-\ref{fig:keyrefresh}
show that if $k=256$, $L \le 128$, and $\rho = 0.5$,
the S-OTS cryptosystem with $q \ge 512$ has sufficient resistance against
the two-stage CPA with probability exceeding $1-10^{-5}$,
regardless of $N$.
Recall from Remark~\ref{rem:tv_bnd} that the S-OTS cryptosystem
must satisfy $q \ge \frac{c_{\rm max}}{4}$ for the indistinguishability,
where $c_{\rm max}$ can be given with respect to $\Psibu$ and $N$.
In the end, we need to carefully choose $q$ by taking into account
both the constraints for indistinguishability and CPA security.

\subsection{Digital Image Encryption}

\begin{figure*}[!t]
\centering{%
\subfloat[Original Lena]{\includegraphics[width=0.15\textwidth, angle=0]{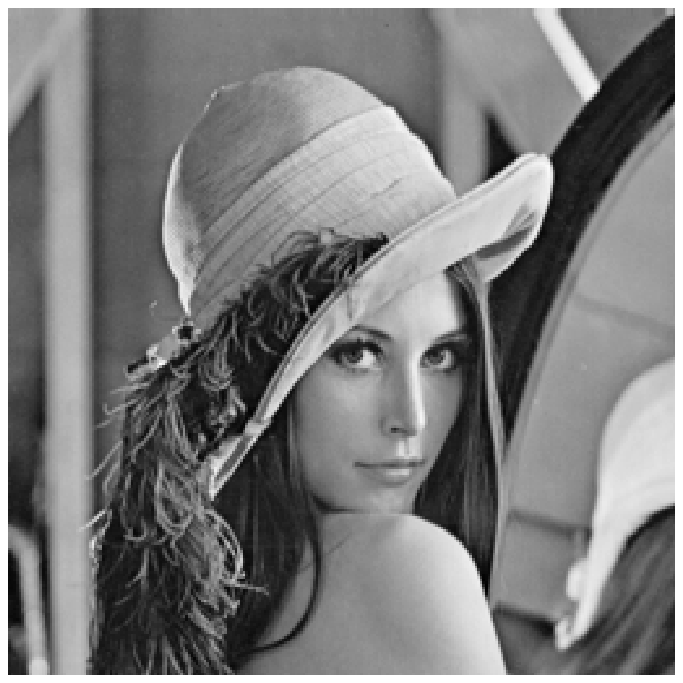}}
\quad
\subfloat[Original Boat]{\includegraphics[width=0.15\textwidth, angle=0]{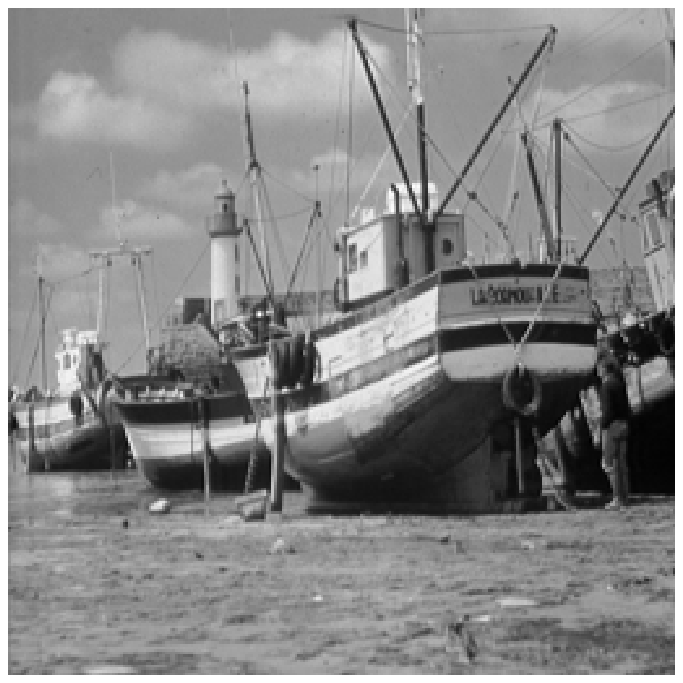}}
\quad
\subfloat[Original Plane]{\includegraphics[width=0.15\textwidth, angle=0]{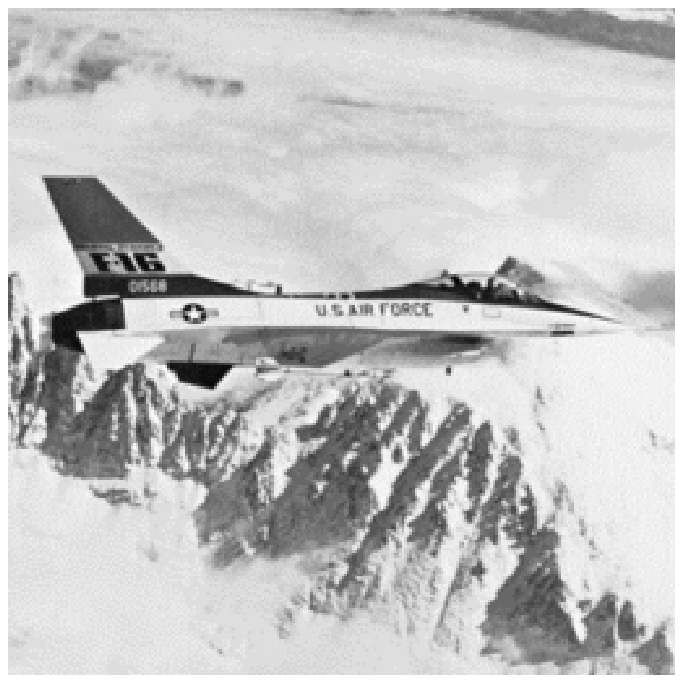}}
\quad
\subfloat[Original Peppers]{\includegraphics[width=0.15\textwidth, angle=0]{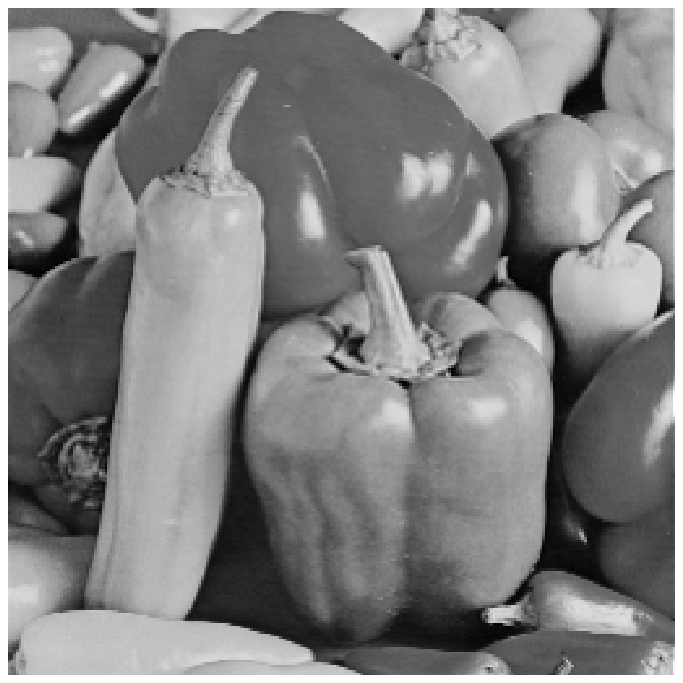}}
\quad
\subfloat[Original Barbara]{\includegraphics[width=0.15\textwidth, angle=0]{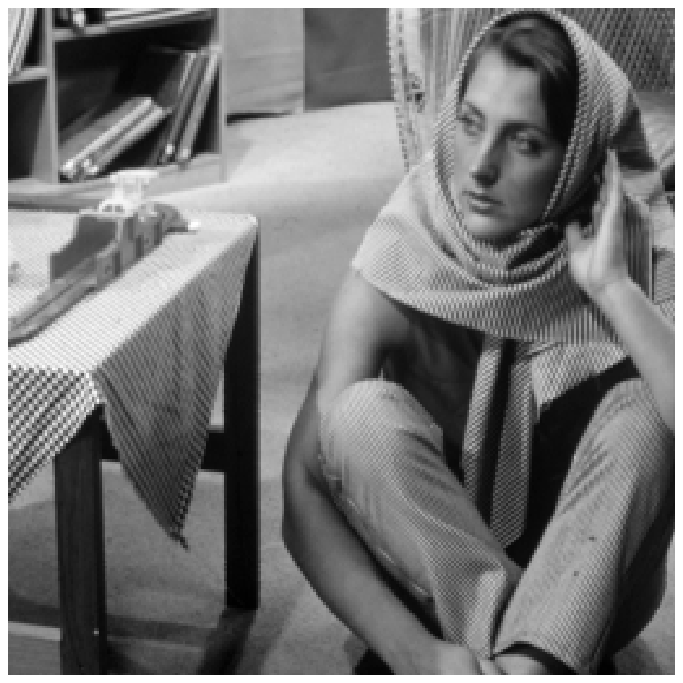}}

\subfloat[Encrypted Lena]{\includegraphics[width=0.15\textwidth, angle=0]{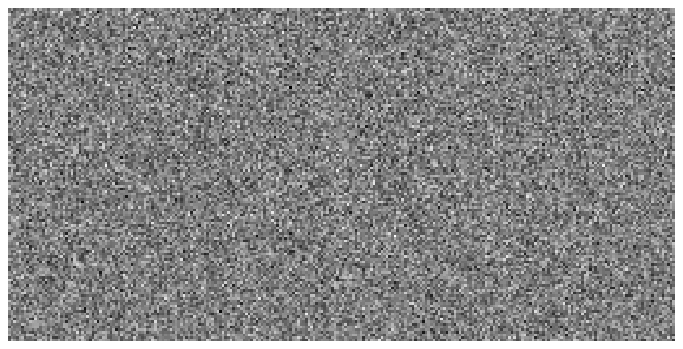}}
\quad
\subfloat[Encrypted Boat]{\includegraphics[width=0.15\textwidth, angle=0]{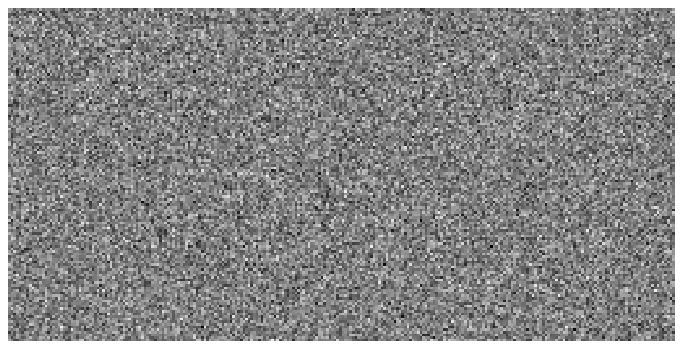}}
\quad
\subfloat[Encrypted Plane]{\includegraphics[width=0.15\textwidth, angle=0]{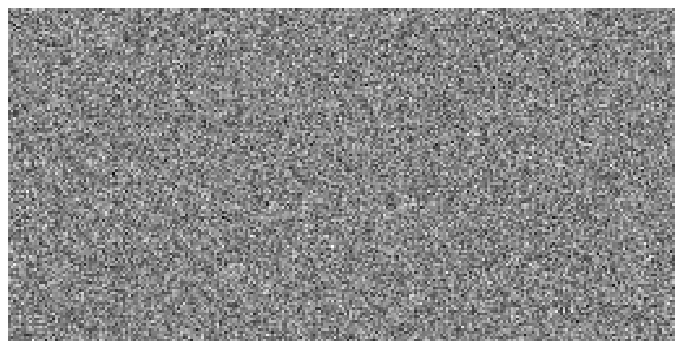}}
\quad
\subfloat[Encrypted Peppers]{\includegraphics[width=0.15\textwidth, angle=0]{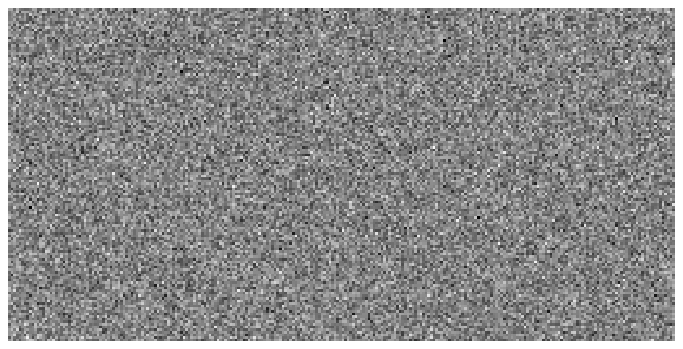}}
\quad
\subfloat[Encrypted Barbara]{\includegraphics[width=0.15\textwidth, angle=0]{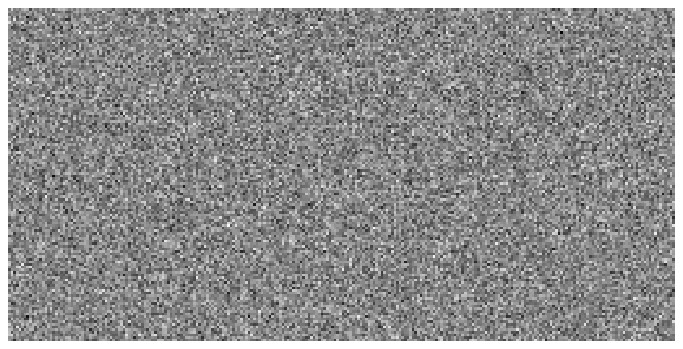}}

\subfloat[Decrypted Lena]{\includegraphics[width=0.15\textwidth, angle=0]{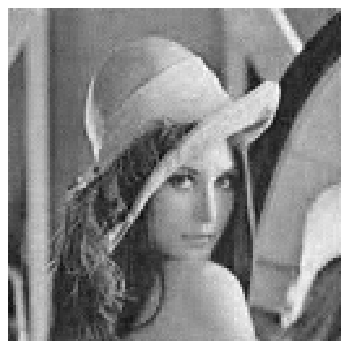}}
\quad
\subfloat[Decrypted Boat]{\includegraphics[width=0.15\textwidth, angle=0]{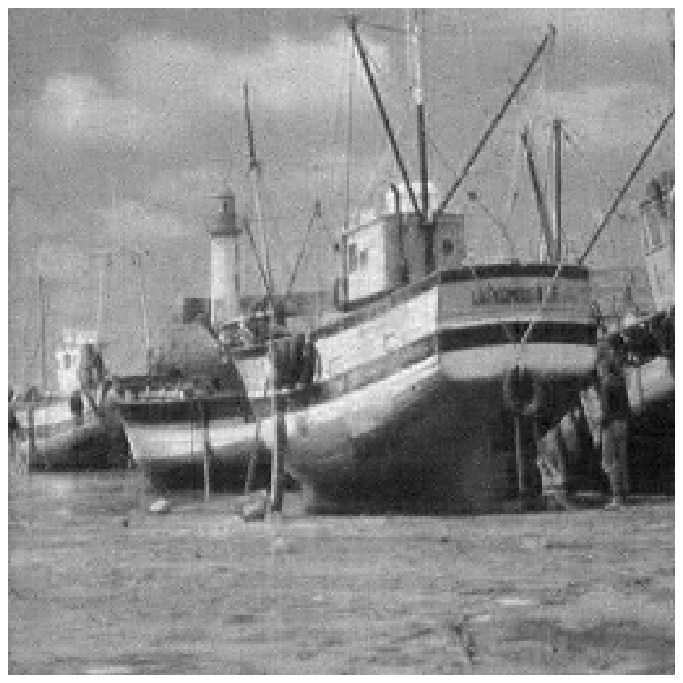}}
\quad
\subfloat[Decrypted Plane]{\includegraphics[width=0.15\textwidth, angle=0]{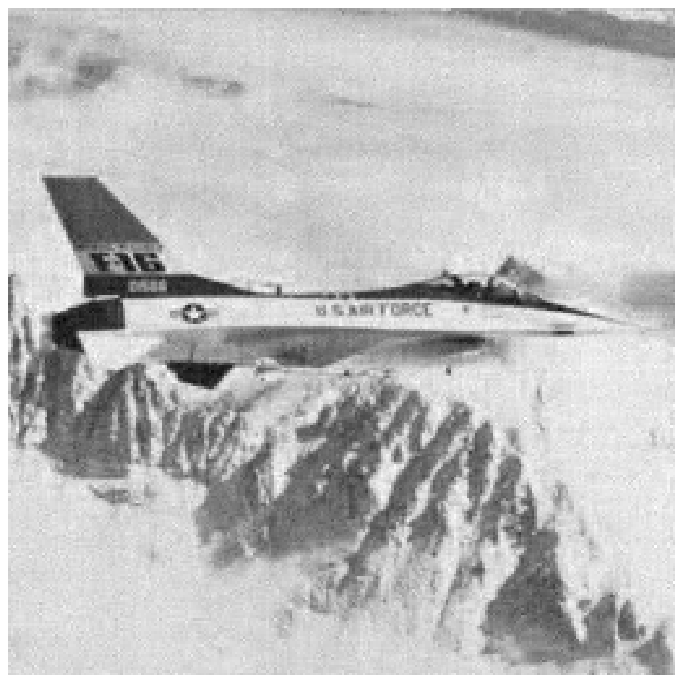}}
\quad
\subfloat[Decrypted Peppers]{\includegraphics[width=0.15\textwidth, angle=0]{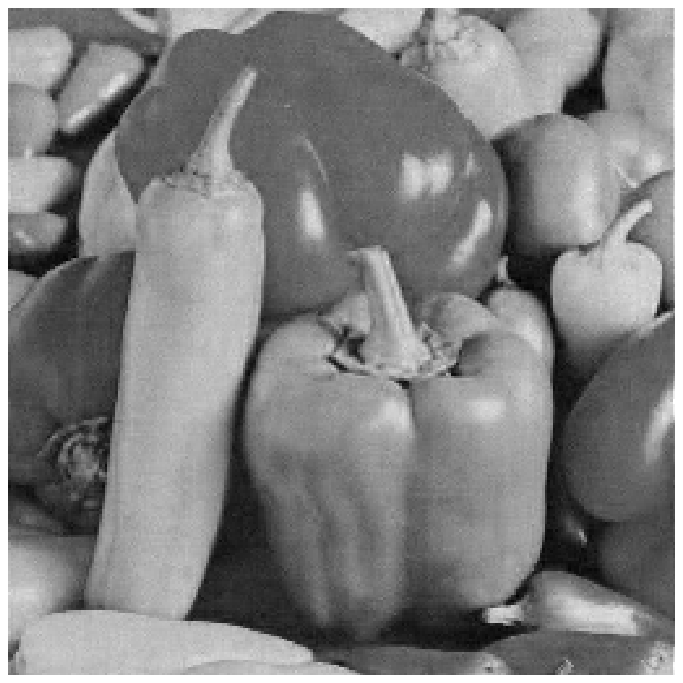}}
\quad
\subfloat[Decrypted Barbara]{\includegraphics[width=0.15\textwidth, angle=0]{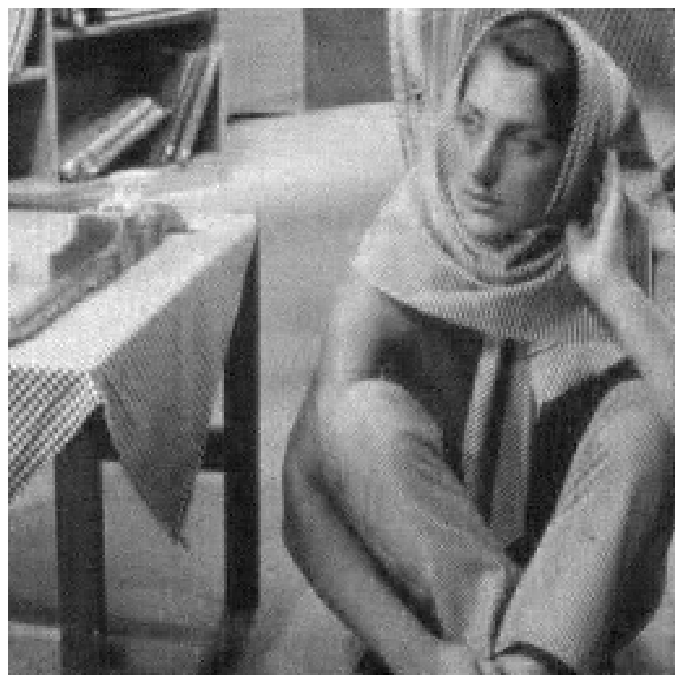}}

}
\caption{Original, encrypted, and decrypted images
of ``Lena'', ``Boat'', ``Plane'', ``Peppers'', and  ``Barbara'', respectively,
where $N=65536$, $q=512$, $k=256$, $\rho=0.5$ and $\Psibu$ is 2D version of the D4 wavelet transform basis.}
\label{fig:image_rec}
\end{figure*}

Recall the simulation setup for digital image encryption in Section II.C.
With $256 \times 256$ images
``Lena'', ``Boat'', ``Plane'', ``Peppers'', and ``Barbara'',
we encrypt each image,
and then decrypt it with SPGL1~\cite{Berg:SPGL1} in noiseless condition, where
2D version of the D4 wavelet transform basis is employed as $\Psibu$.
We select $q=512$, which meets the requirements for indistinguishability and CPA security.
Figure~\ref{fig:image_rec} visualizes the
original, encrypted, and decrypted images, respectively,
where $\rho = 0.5$.
The encrypted images are visually unrecognizable
and then successfully decrypted with the knowledge of $\Phibu$,
where PSNR values are $32.2$ dB, $29.6$ dB $31.1$ dB, $31.6$ dB, and $29.5$ dB, respectively.
In the examples of Figure~\ref{fig:image_rec},
the measurement matrix $\Phibu$ is highly sparse with the row-wise sparsity
$r = \frac{q}{N} = 2^{-7} \approx 0.78\%$.
Using the sparse matrix $\Phibu$,
the S-OTS cryptosystem requires $c_s + c_p \approx 2^{24}$
bits of the SSG output sequence for each encryption process,
while the B-OTS cryptosystem uses $MN=2^{31}$ bits.
This implies that the S-OTS cryptosystem can save a significant amount of
keystream bits while guaranteeing security against the two-stage CPA.

\section{Conclusion}

In this paper, we proposed the S-OTS cryptosystem,
which employs sparse measurement matrices
for secure and efficient CS encryption.
With a small number of nonzero elements in the measurement matrix,
the S-OTS cryptosystem has complexity benefits in terms of
memory and computing resources.
In addition, the S-OTS cryptosystem can present theoretically
guaranteed CS recovery performance for a legitimate recipient.
In the presence of an adversary,
we analyzed the security of the S-OTS cryptosystem against COA and CPA.
Against COA, we exhibited that the S-OTS cryptosystem can asymptotically achieve
the indistinguishability, as long as each plaintext has constant energy.
To investigate its security against CPA, we consider an adversary's strategy that
consists of two sequential stages, keystream and key recovery attacks.
We then showed that the keystream recovery
can be infeasible with overwhelmingly high probability.
Also, we conducted an information-theoretic analysis to
demonstrate that the success probability of following key recovery
can be extremely low with a proper selection of parameters.
Through numerical results,
we demonstrated that the S-OTS cryptosystem guarantees
its reliability and security, while providing computational efficiency.

\appendix

\subsection{Example of Permutation Recovery Attack}~\label{pf:permut_keystream}

In Section IV.B, a keystream recovery attack with the second class plaintext can provide an adversary with the positions of nonzero entries in $\Phibu$, which can be exploited to recover $\Pbu$. 
Let $\Lambda_i^{\Phibu}$ be an index set of nonzero entries in the $i$-th row of $\Phibu$.
Given $\Lambda_i^{\Phibu}$ for $i=1,\cdots,M$,
an adversary may attempt a known plaintext attack against the \emph{permutation-only} cipher~\cite{Zhang:permut} to find a true $\Pbu$,
where a plaintext ${\bf p}_i = (p_{i,1},\cdots,p_{i,N})$
and the corresponding ciphertext ${\bf c}_i= (c_{i,1},\cdots,c_{i,N})$
with respect to $\Lambda_i$ and $\Lambda_i^{\Phibu}$ are given by
\begin{equation*}
    p_{i,j} =
    \begin{cases}
    1, & \text{if } j \in \Lambda_i, \\
    0, & \text{otherwise},
    \end{cases}
    \quad
    \text{and}\quad
    c_{i,j} =
    \begin{cases}
    1, & \text{if } j \in \Lambda^{\Phibu}_{i}, \\
    0, & \text{otherwise}.
    \end{cases}
\end{equation*}
Let ${\bf C} \in \{0,1\}^{M\times N}$ be a matrix
having ${\bf c}_i$ as its $i$-th row for $i=1,\cdots,M$ and
$\bar{\bf c} \in \R^N$ be a \emph{composite representation}~\cite{Zhang:permut}
of ${\bf C}$, where the $j$-th element of $\bar{\bf c}$ is given by
\begin{equation*}\label{eq:comp_rep}
\bar{c}_j = \sum_{i=1}^M c_{i,j} \cdot 2^{i-1}.
\end{equation*}
According to~\cite[Proposition 2]{Zhang:permut}, the permutation pattern
can be uniquely determined if and only if all the entries of $\bar{\bf c}$ are distinct.
In the S-OTS cryptosystem, however,
$\bar{\bf c}$ consists of $\eta = \frac{N}{q}$ distinct integers $\{z_1,\cdots,z_\eta\}$,
where $z_i$ appears $q$ times in $\bar{\bf c}$
for all $i=1,\cdots,\eta$, due to the structure of $\Sbu$ defined in~\eqref{eq:set_define} and~\eqref{eq:s_structure}.
Then, the number of possible permutation patterns is
\begin{equation*}\label{eq:s_p}
  \mathcal{S}_{\Pbu} = (q!)^\eta,
\end{equation*}
which is smaller than $N!$, but still extremely large even for
small $q$ and $N$.
Although we can reduce the number of possible permutations by 
observing the nonzero entries of $\Phibu$, retrieving $\Pbu$
can be intractable to an adversary with bounded computing power.
Note that this example does not take into account the potential of
a more elaborate attack exploiting the respective positions of
$+1$'s and $-1$'s in $\Phibu$, which
is left open for future research.

\subsection{Proof of Theorem~\ref{th:cpa_sol}}~\label{pf:cpa_sol}
Without loss of generality, let $\Gamma = \{1,\cdots,\tau\}$ be a set
of the indices of the first $\tau$ rows in $\Sbu$.
Then, an adversary can obtain $\widehat{\bf b}^k$
by estimating the $i$-th row of $\Sbu$, or $\sbu^{(i)}$, for every $i \in \Gamma$.
Given $q_i^+ = \frac{1}{2}(q+y_i)$
and $q_i^- = \frac{1}{2}(q-y_i)$ from~\eqref{eq:ML_model},
the number of possible solutions for $\sbu^{(i)}$ is given by
\begin{equation}\label{eq:s_i}
\mathcal{S}_i = \binom{q}{q_i^+}=\binom{q}{q_i^-}.
\end{equation}

Assuming that each nonzero entry of $\Sbu$ takes $\pm 1$ independently with probability $0.5$,
the sum of $q$ independent random variables yields
$y_i=q_i^+ - q_i^- = 2q_i^+ -q  \in \{-q,\cdots,q\}$,
which can be considered as a binomial random variable.
Therefore, the Hoeffding's inequality~\cite{Hoeffding:ineq} yields
\begin{equation}\label{eq:hoeffding}
  {\rm Pr}\left[\left|y_i\right|<t\right] = {\rm Pr}\left[\frac{q-t}{2}<q_i^+<\frac{q+t}{2}\right]
  \ge 1-2e^{\frac{-t^2}{2q}}.
\end{equation}
Since $q_i^+$ takes an integer value,
\begin{equation*}
  {\rm Pr}\left[\frac{q-t}{2}<q_i^+<\frac{q+t}{2}\right]
  = {\rm Pr}\left[\left\lceil\frac{q-t}{2}\right\rceil\le q_i^+ \le \left\lfloor\frac{q+t}{2}\right\rfloor\right].
\end{equation*}
From~\eqref{eq:s_i} and~\eqref{eq:hoeffding}, $\mathcal{S}_i$ can be bounded by
\begin{equation*}
  \mathcal{S}_i = \binom{q}{q_i^+} \ge \binom{q}{\left\lceil\frac{q-t}{2}\right\rceil},
\end{equation*}
with probability exceeding $1-2e^{\frac{-t^2}{2q}}$.
Finally, the number of all possible solutions for $\tau$ consecutive rows is
$\mathcal{S}_{\rm CPA}=\prod_{i \in \Gamma} \mathcal{S}_i$, where
\begin{equation}\label{eq:cpa_prob}
  \begin{split}
  {\rm Pr}\left[\mathcal{S}_{\rm CPA}\ge\binom{q}{\left\lceil\frac{q-t}{2}\right\rceil}^\tau\right]
  &\ge \prod_{i \in \Gamma}{\rm Pr}\left[\mathcal{S}_i \ge\binom{q}{\left\lceil\frac{q-t}{2}\right\rceil}\right]\\
  &\ge \left(1-2e^{\frac{-t^2}{2q}}\right)^{\tau}.
\end{split}
\end{equation}
Letting $\left(1-2e^{\frac{-t^2}{2q}}\right)^{\tau} = 1-\varepsilon_2$,
$t=\sqrt{2q \cdot \log \frac{2}{1-(1-\varepsilon_2)^{\frac{1}{\tau}}}}$,
which completes the proof.
\qed

\subsection{Proof of Theorem~\ref{th:cpa_q}}~\label{pf:CPA_q}
In~\eqref{eq:CPA_sol}, let $\lceil\frac{q-t}{2}\rceil = \alpha$.
Then, $\tau = \lceil \frac{k}{q} \rceil \ge \frac{k}{q}$ yields
\begin{equation*}
  \mathcal{S}_{\rm CPA} \ge \binom{q}{\alpha}^\tau \ge \binom{q}{\alpha}^{\frac{k}{q}}
  > \left(\frac{q}{\alpha}\right)^{\alpha \cdot\frac{k}{q}}.
\end{equation*}
Therefore, $\mathcal{S}_{\rm CPA} > 2^L$
if $q$ satisfies $(\frac{q}{\alpha})^{\alpha \cdot\frac{k}{q}} \ge 2^L$,
which is equivalent to
\begin{equation}\label{eq:CPA_q_eq1}
\frac{\alpha}{q} \log \frac{\alpha}{q} \le - \frac{L \log 2}{k}.
\end{equation}
Since $\frac{1}{x} \log \frac{1}{x} \ge -\frac{1}{e}$ for $x>0$,
\eqref{eq:CPA_q_eq1} is valid as long as
\begin{equation}\label{eq:key_cond}
k \ge L \cdot e \log 2.
\end{equation}
By taking ${\mathcal W}_{-1}(\cdot)$,~\eqref{eq:CPA_q_eq1} yields
$q \le \alpha \beta$,
where $\alpha =\lceil\frac{q-t}{2}\rceil$, $t=\sqrt{2q \cdot \log \frac{2}{1-(1-\varepsilon_2)^{\frac{1}{\tau}}}}$,
and $\beta =  -\frac{k}{L \log 2} {\mathcal W}_{-1}\left(-\frac{L \log 2}{k}\right)$.
From $q \le \alpha \beta$, we have
\begin{equation}\label{eq:CPA_q_eq4}
  q \ge \frac{1}{2} \left(2+\frac{4}{\beta - 2}\right)^2 \log \frac{2}{1-(1-\varepsilon_2)^{\frac{1}{\tau}}},
\end{equation}
which is a sufficient condition for $\mathcal{S}_{\rm CPA} > 2^L$.
Furthermore,
since $\tau = \lceil \frac{k}{q} \rceil < \frac{kM}{N} +1$ from $\frac{N}{M} \le q$,
we have
$1-(1-\varepsilon_2)^{\frac{1}{\tau}} > 1-(1-\varepsilon_2)^{\frac{1}{k \rho +1}}$,
where $\rho = \frac{M}{N}$ and the sufficient condition of~\eqref{eq:CPA_q_eq4} becomes
\eqref{eq:CPA_p},
which completes the proof.
\qed

\subsection{Proof of Theorem~\ref{th:key_prob}}~\label{pf:key_prob}
Based on $\widehat{\bf b}^k$,
an adversary attempts to recover the true key ${\bf k}$
by choosing a candidate key $\widehat{\bf k}$.
Let ${\rm P}_{\rm err}$ be the error probability of key recovery, where
\begin{equation}\label{eq:fano1}
  \begin{split}
{\rm P}_{\rm err}
&= {\rm Pr}\left[\widehat{\bf k} \neq {\bf k} | \widehat{\bf b}^k = {\bf b}^k \right] \cdot {\rm Pr}\left[\widehat{\bf b}^k = {\bf b}^k \right]\\
&\quad+ {\rm Pr}\left[\widehat{\bf k} \neq {\bf k} | \widehat{\bf b}^k \neq {\bf b}^k \right] \cdot {\rm Pr}\left[\widehat{\bf b}^k \neq {\bf b}^k \right].
\end{split}
\end{equation}
When $\widehat{\bf b}^k ={\bf b}^k$,
the data processing and the Fano's inequalities~\cite{Cover:inf} yield
\begin{equation}\label{eq:fano2}
H \left({\bf k} \arrowvert {\bf b}^k \right)  \le H \left({\bf k} \arrowvert \widehat{\bf k} \right)
 \le H_b(p_e) + p_e \cdot \log_2 \left|\mathcal{K}\right|,
\end{equation}
where $H(\cdot)$ is the entropy of a random vector,
$p_e={\rm Pr}[\widehat{\bf k} \neq {\bf k} | \widehat{\bf b}^k = {\bf b}^k]$,
$H_b(p_e) = - p_e \log_2 p_e -(1-p_e) \log_2 (1-p_e)$, and
$\mathcal{K}$ is a set of all possible candidates of ${\bf k}$, i.e., $|\mathcal{K}| = 2^k$.
Given ${\bf b}^k$,
assume that $2^{\delta k}$ candidates of ${\bf k}$ are uniformly distributed,
i.e., $H \left({\bf k} \arrowvert {\bf b}^k \right) = \delta k$.
Using $H_b(p_e) \le 1$,~\eqref{eq:fano2} yields
\begin{equation}\label{eq:fano3}
p_e = {\rm Pr}\left[\widehat{\bf k} \neq {\bf k} | \widehat{\bf b}^k = {\bf b}^k \right] \ge \delta - \frac{1}{k}.
\end{equation}
When the adversary has a wrong keystream,
i.e., $\widehat{\bf b}^k \neq {\bf b}^k$,
we assume that the success probability of key recovery becomes
\begin{equation}\label{eq:brute}
 {\rm Pr}\left[\widehat{\bf k} = {\bf k} | \widehat{\bf b}^k \neq  {\bf b}^k\right] = 2^{-k}.
\end{equation}
With~\eqref{eq:fano3},~\eqref{eq:brute}, and ${\rm P}_{\rm suc} = {\rm Pr}[\widehat{\bf b}^k = {\bf b}^k]$,
\eqref{eq:fano1} yields
\begin{equation*}\label{eq:fano5}
  \begin{split}
{\rm P}_{\rm err}
&\ge \left(\delta - \frac{1}{k} \right) \cdot {\rm P}_{\rm suc} + (1-2^{-k}) \cdot \left(1-{\rm P}_{\rm suc}\right)\\
&\ge 1-2^{-k}-\left(1-2^{-k}-\delta + \frac{1}{k}\right) \cdot {\rm P}_{\rm suc,up},
\end{split}
\end{equation*}
which completes the proof by ${\rm P}_{\rm key} = 1 - {\rm P}_{\rm err}$.
\qed

\end{document}